\documentclass[fleqn,usenatbib]{mnras}

\usepackage{newtxtext,newtxmath}

\usepackage[T1]{fontenc}

\DeclareRobustCommand{\VAN}[3]{#2}
\let\VANthebibliography\thebibliography
\def\thebibliography{\DeclareRobustCommand{\VAN}[3]{##3}\VANthebibliography}

\usepackage{graphicx}	
\usepackage{amsmath}	
\usepackage{enumitem}
\usepackage{array}

\newcommand{\absval}[1]{\left\lvert#1\right\rvert}
\newcommand{\norm}[1]{\left\lVert#1\right\rVert}

\title[REACH calibrator set optimisation]{Optimisation of calibration sources for global 21-cm experiments: the REACH case}

\author[A.K. Dash et al.]{
Adarsh Kumar Dash\textsuperscript{\href{https://orcid.org/0000-0002-7544-7881}{\includegraphics[width=2.5mm]{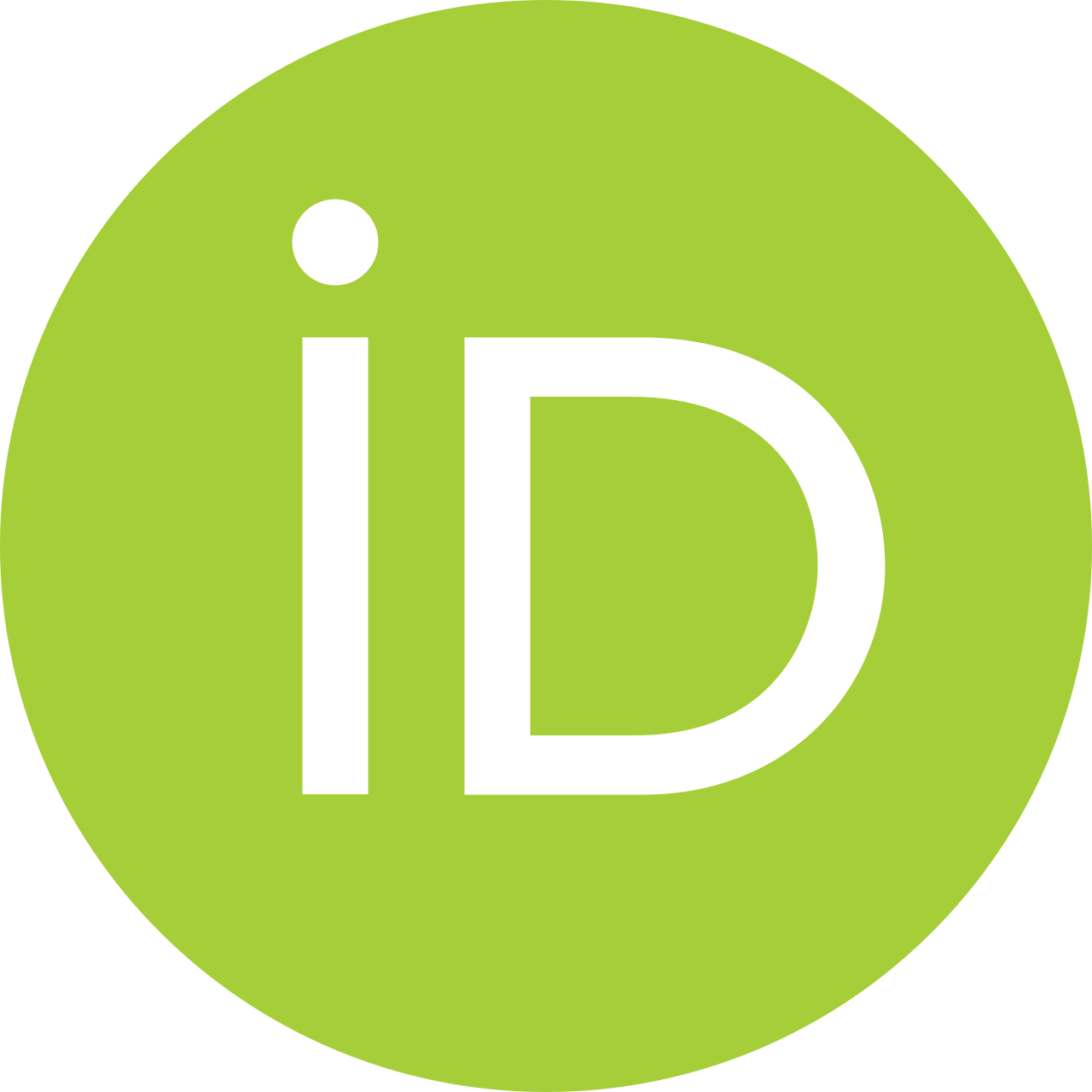}}},$^{1,2}$\thanks{E-mail: adarsh242k@gmail.com}
Dominic Anstey,$^{1,2}$
Harry T. J. Bevins,$^{1,2}$
Eloy de Lera Acedo,$^{1,2}$
\newauthor
Gary Allen,$^{4}$
Kaan Artuc,$^{1,2}$
Gianni Bernardi,$^{7}$
Martin Bucher,$^{4,6}$
Steve Carey,$^{1}$
Jean Cavillot,$^{8}$
\newauthor
Ricardo Chiello,$^{9}$
Adelicia S. Chu,$^{1,2}$
Wessel Croukamp,$^{4}$
John Cumner,$^{1,2}$
Saswata Dasgupta,$^{2,3}$
\newauthor
Dirk I. L. de Villiers,$^{4}$
Jiten Dhandha,$^{2,3}$
Aleksandra Dragovic,$^{1,2}$
John A. Ely,$^{1}$
Anastasia Fialkov,$^{2,3}$
\newauthor
Thomas Gessey-Jones,$^{1,2}$
Will J. Handley,$^{2,3}$
Christian Kirkham\textsuperscript{\href{https://orcid.org/0000-0001-5385-6329}{\includegraphics[width=2.5mm]{ORCID_iD.svg.png}}},$^{1,2}$
Girish Kulkarni,$^{10}$
\newauthor
Samuel A. K. Leeney,$^{1,2}$
Alessio Magro,$^{5}$
P. Daan Meerburg,$^{11}$
Shikhar Mittal\textsuperscript{\href{https://orcid.org/0000-0002-0247-618X}{\includegraphics[width=2.5mm]{ORCID_iD.svg.png}}},$^{1,2}$
Daniel Molnar,$^{1,2}$
\newauthor
Rohan S. Patel,$^{1,2}$
Joe H. N. Pattison,$^{1,2}$
Saurabh Pegwal,$^{4}$
Carla M. Pieterse,$^{4}$
Jonathan R. Pritchard,$^{12}$
\newauthor
Gabriella Rajpoot,$^{1,2}$
Nima Razavi-Ghods,$^{1}$
Daniel Robins,$^{1,2}$
Ian L. V. Roque,$^{1}$
Anchal Saxena,$^{1,2}$
\newauthor
Killian H. Scheutwinkel,$^{1,2}$
Emma Shen,$^{1,2}$
Peter H. Sims,$^{1,2}$
Marta Spinelli,$^{13,14}$ 
Jiacong Zhu$^{1,15}$
\\
$^{1}$Astrophysics Group, Cavendish Laboratory, University of Cambridge, J. J. Thomson Avenue, Cambridge, CB3 0US, UK\\
$^{2}$Kavli Institute for Cosmology in Cambridge, University of Cambridge, Madingley Road, Cambridge, CB3 0HA, UK\\
$^{3}$Institute of Astronomy, University of Cambridge, Madingley Road, Cambridge, CB3 0HA, UK\\
$^{4}$Department of Electrical and Electronic Engineering, Stellenbosch University, Stellenbosch, 7602, South Africa\\
$^{5}$Institute of Space Sciences and Astronomy, University of Malta, Msida, Malta, MSD 2080, Malta\\
$^{6}$Laboratoire AstroParticule et Cosmologie, Université Paris-Cité, 10 Rue Alice Domon et Léonie Duquet, Paris, 75013, France\\
$^{7}$INAF-Istituto di Radio Astronomia, Via Gobetti 101, Bologna, 40129, Italy\\
$^{8}$Antenna Group, Université catholique de Louvain, Louvain-la-Neuve, 1348, Belgium\\
$^{9}$Physics Department, University of Oxford, Parks Road, Oxford, OX1 3PU, UK\\
$^{10}$Department of Theoretical Physics, Tata Institute of Fundamental Research, Homi Bhabha Road, Mumbai, 400005, India\\
$^{11}$Faculty of Science and Engineering, University of Groningen, Nijenborgh 4, Groningen, 9747 AG, Netherlands\\
$^{12}$Max Planck Institute for Radio Astronomy, Auf dem Hügel 69, 53121 Bonn, Germany\\
$^{13}$Observatoire de la Côte d'Azur, Nice, France\\
$^{14}$Department of Physics and Astronomy, University of the Western Cape, Robert Sobukhwe Road, Bellville, 7535, South Africa\\
$^{15}$National Astronomical Observatory, Chinese Academy of Sciences, Beijing, 100101, China
}

\date{Accepted XXX. Received YYY; in original form ZZZ}

\pubyear{\the\year{}}

\begin{document}
\label{firstpage}
\pagerange{\pageref{firstpage}--\pageref{lastpage}}
\maketitle



\begin{abstract}
The spin-flip 21-cm signal from the Cosmic Dawn and the Epoch of Reionization is an essential probe of the conditions that led to the formation of the first luminous objects in the early Universe. However, its detection remains a major challenge owing to its low strength compared to the bright foregrounds and the requirement of precise calibration of the instrument to prevent systematics that could hinder a detection or lead to false inferences.
REACH (Radio Experiment for the Analysis of Cosmic Hydrogen) is a radiometer experiment designed to detect this sky-averaged signal in the frequency range of 50--130~MHz. Using a wide-beam antenna, REACH calibration relies on internal reference sources, covering a broad range of temperatures and reflection coefficients. 
The choice of type and number of calibrators used significantly influences the quality of the calibration. This work investigates these effects and introduces a novel method for selecting an optimal set of calibration sources. With an optimised set, we aim to reduce calibration time, thereby increasing sky integration time while preserving calibration accuracy.
We explore two optimisation strategies: one applied across the full receiver band and another performed on a frequency-by-frequency basis. 
Finally, we demonstrate that, with a total calibration time comparable to the conventional full-calibrator set, an optimised set with fewer calibrators achieves approximately a $15~\%$ reduction in calibrated temperature noise and improved absolute calibration of the instrument. This has implications for better calibration strategies in similar radiometer experiments.
\end{abstract}

\begin{keywords}
instrumentation: interferometers -- methods: data analysis -- dark ages, reionization, first stars -- early Universe -- cosmology: observations
\end{keywords}


\section{Introduction}
The Cosmic Dawn and Epoch of Reionization mark the period of evolution of the first stars and galaxies, transitioning the Universe from a neutral to a nearly ionised state \citep{Shaver_1999, Furlanetto_2004, Furlanetto_2006, 21cm_cosmology_Pritchard_Loeb}. Studying this period can provide a rich understanding of the astrophysical and cosmological processes in the early Universe \citep{EoR_Global_21cm}. A useful probe to study this period, spanning the redshifts $z~\sim~6$--$35$ \citep{Cohen_21cm_parameter_space}, is the spin-flip 21-cm signal emitted by the neutral Hydrogen (\ion{H}{I}) present in the early Universe \citep{Madau_1997}.

This signal is predicted to have redshifted from its rest emission frequency of $\nu~\sim~1420.4$~MHz to the range 40--200~MHz, with a maximum absolute brightness temperature of a few hundred mK \citep{Cohen_21cm_parameter_space, Mittal_ECHO21}. A major challenge for the detection of the 21-cm signal is the bright galactic and extra-galactic foregrounds with intensities of the order $10^3$--$10^4$~K \citep{GSM_de_Oliveira, GMOSS_Rao_2017, Mittal_point_sources_impact}. Moreover, the observation band is also contaminated by various terrestrial and satellite sources of radio frequency interference (RFI) \citep{Offringa_RFI_LOFAR, HERA_RFI, Bayesian_RFI_flagging}. This demands an accurate and precise calibration of any radiometer aimed at detecting this faint signal. 

There are several experiments designed to detect the sky-averaged component of this signal, such as REACH \citep{REACH_paper, REACH_radiometer_design}, EDGES \citep{Monsalve_2017_EDGES_receiver_calibration, EDGESpaper}, SARAS \citep{SARAS3paper}, EIGSEP \citep{EIGSEP_paper}, $\mathrm{PRI^ZM}$ \citep{PRIZM_paper}, RHINO \citep{RHINOpaper} and MIST \citep{MISTpaper}. 
In addition to these ground-based radiometers, experiments like the CosmoCube \citep{CosmoCube_spectrometer, CosmoCube_receiver_calibration}, PRATUSH \citep{PRATUSHpaper}, DARE \citep{DAREpaper, DARE_observation_strategy}, and DAPPER \citep{DAPPERpaper} are a few radiometers which are being developed for observing the 21-cm signal from space, opening a new observational domain for global 21-cm cosmology. 
In 2018, the EDGES experiment claimed the first detection of this signal centred at 78~MHz, and modelled it as a flat Gaussian \citep{EDGESpaper}. With an amplitude of $\sim$~500~mK, this feature was much deeper than that predicted by the standard $\Lambda$CDM cosmology \citep{Cohen_21cm_parameter_space}. 
While this observation could be explained by introducing excess cooling of interstellar gas due to its interaction with dark matter \citep{Interstellar_gas_cooling_edges} or an excess radio background \citep{excess_radio_edges, Excess_radio_Ewall}, several studies have raised concerns about the interpretation of this detection, suggesting the presence of an uncorrected systematic in the data \citep{Hills2018_concerns_EDGESmodel, Bradley_2019_EDGESmodel, Singh_2019_EDGESmodel, Sims_and_Prober_EDGESmodel_test,  Harry_maxsmooth_EDGES}. 
Later in 2022, the SARAS experiment reached a similar noise level as that of EDGES and refuted this detection with 95.3\% confidence \citep{SARAS3_constraints_measurement}.

In this work, we focus on the REACH radiometer (Radio Experiment for the Analysis of Cosmic Hydrogen), primarily on the calibration of its receiver. The radiometer uses a wide-beam fixed hexagonal dipole antenna, covering the frequency range 50--130~MHz, and is located in the Karoo radio reserve in South Africa \citep{REACH_antenna}.
The use of a fixed wide-beam antenna in REACH makes the conventional calibration methods impractical. To correct the response of the complete receiver, these methods typically involve observing a known external source, usually a point source, which cannot be targetted or resolved in this case.
This necessitates the use of internal reference sources with known characteristics.
The calibration process, therefore, aims to recover the true brightness of an observed source by correcting for the instrumental responses derived by observing these reference sources. 

For calibration, REACH uses temperature and impedance (reflection coefficient) information from a diverse set of such sources (calibrators). 
This information is used to derive a noise-wave formalism, where the noise power produced by the sources is treated as voltage waves \citep{Meys_noise_wave}, yielding noise-wave parameters for the receiver that are estimated during the calibration process. REACH primarily implements two methods for this estimation: linear least-squares \citep{REACH_radiometer_design} and a Bayesian framework, in which the prior and posterior are conjugate distributions, giving a closed form expression for the posterior \citep{Bayesian_nwp_conjugate_priors, kirkham2025accountingnoisesingularitiesbayesian}. Additionally, a machine learning-based calibration framework has also been developed, which uses a noise parameter formalism \citep{Leeney2025_ML}. 

Here, we explicitly use the least-squares method and relate the properties of the internal calibration sources to a metric, namely the \textit{condition number} \citep{Condition_number}. This is used to put a preference on the sources to be used in the calibrator set. We study the effects of this `preference' on the calibration accuracy. 
Further motivation for this optimisation stems from the fact that every REACH measurement configuration dedicates a significant amount of time to calibration, observing the 12 calibration sources. We expect the information from a few of these sources to be degenerate at certain frequencies, and thus an optimised set would reduce the calibration period, providing more time for sky observation. 
Moreover, reducing the total duration of measurement could, in principle, reduce any drift errors that could occur in a complete measurement set \citep{kirkham2025capturingdrifttimeseries}.

The paper is divided into four main sections. We start with an overview of the REACH calibration in Section~\ref{sec:REACH_calibration_overview}, where we introduce the calibration equation and methodology, the receiver design, and the methods of noise estimation for REACH, useful for analysing the calibration results later in the paper.
This section also introduces condition numbers as a function of the receiver design matrix $\mathbf{X}$ in the least-squares formalism, which is defined based on the calibration sources. 
Section~\ref{sec:Application_to_reach_data} argues why the condition number is a useful metric to ascertain the `goodness' of calibration. This leads to the method of optimisation of the calibrator set and we present the optimised set for REACH receiver calibration.
We also discuss another method, namely piecewise calibration, where the said calibrator set optimisation is performed for each frequency channel, instead of the complete frequency band.
Using the approaches of optimisation, we perform calibration on REACH measurements, and compare its performance with the original full calibrator set and discuss the results in Section~\ref{sec:calibration_results}.
Finally, we summarise the paper in Section~\ref{sec:conclusion}.

\section{REACH receiver calibration overview} \label{sec:REACH_calibration_overview}
 
In this section, we describe the method of calibration for the REACH receiver.
Following a calibration for a source, it is important to validate the solution against a reference. For this purpose, we introduce two ways of noise estimation later in this section.
Finally, we present the concept of condition numbers, which shall be used as a key tool for choice of calibration sources in this paper. 

\subsection{The calibration formalism}
 
\begin{figure}
    \centering
    \includegraphics[width=1\linewidth]{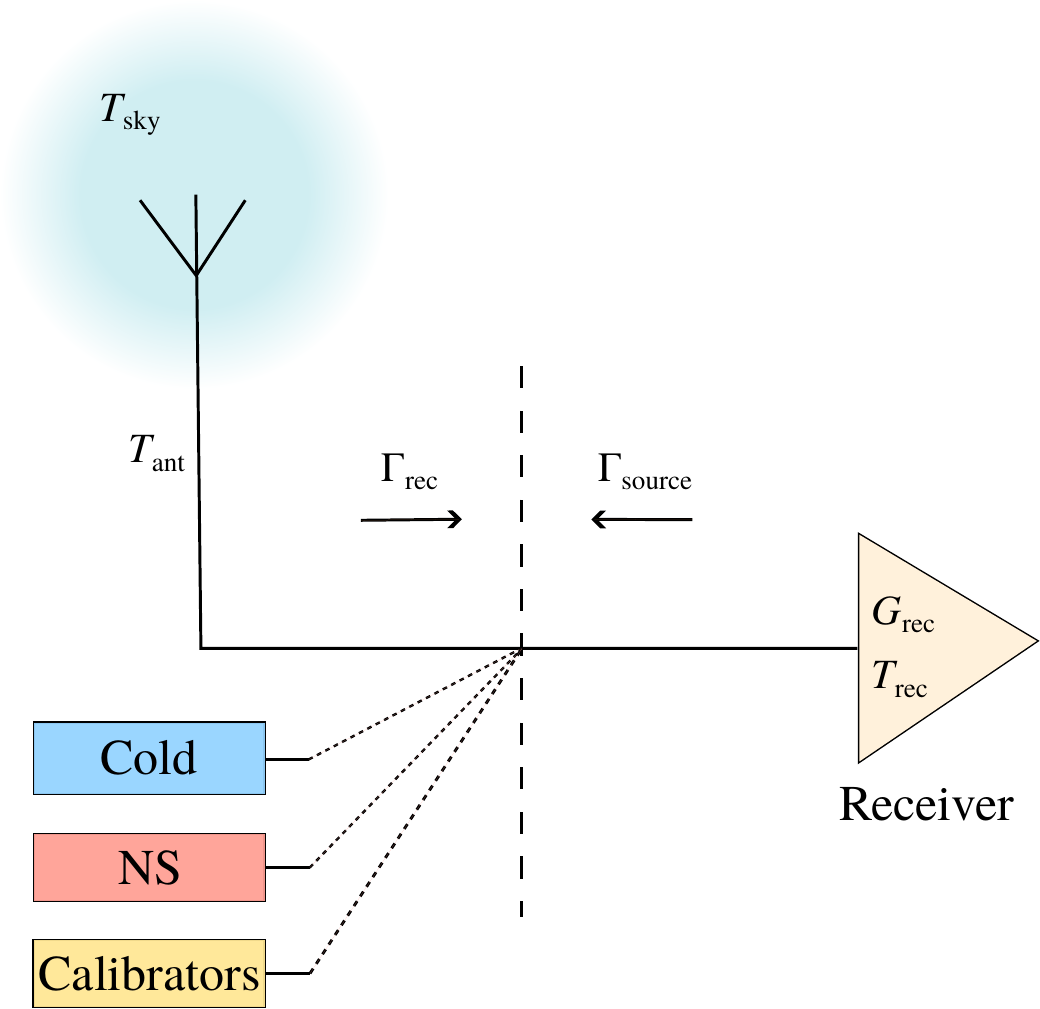}
    \caption{A schematic of a simple radiometer. The antenna observes the sky at brightness temperature $T_\mathrm{sky}$. $T_\mathrm{ant}$ is the antenna temperature. The long dashed line represents the reference plane where the impedance mismatch is supposed to occur between a source and the receiver. $G_\mathrm{rec}$ and $G_\mathrm{rec}$ are the gain and noise temperature of the receiver. The calibration is performed by Dicke switching between the cold source and noise source (NS), as well as switching among the calibrators in the case of a mismatched antenna.}  
    \label{fig:Simple_receiver}
\end{figure}

The instrument responses and systematic errors that we aim to correct in receiver calibration include the gain and additive noise from the receiver, and reflections between the various elements in the signal path arising from impedance mismatches.
These are illustrated in a simple schematic in Figure~\ref{fig:Simple_receiver}, which shows the signal chain of a radiometer. The signal path consists of an antenna of reflection coefficient $\Gamma_\mathrm{ant}$ connected to the receiver with reflection coefficient $\Gamma_\mathrm{rec}$. 
Without loss of generality, the receiver is sometimes denoted as an LNA. The antenna is observing the sky at a temperature $T_\mathrm{sky}$, which gets coupled with the antenna's beam pattern, and further modulated by its radiation efficiency and reflection coefficient to give the antenna temperature $T_\mathrm{ant}$ \citep{Balanis_Antenna_theory}. 
If both the antenna and receiver are perfectly matched, i.e. the impedances $Z_\mathrm{ant} = Z_\mathrm{rec} (= Z_\mathrm{ref})$, the reflections at each port would become zero. Here, we consider the reference impedance $Z_\mathrm{ref} = 50~\Omega$. However, due to limitations in design capabilities, this is not usually true, resulting in mismatches and reflections. 

We start the derivation of the calibration equation with the assumption of a perfectly matched source at temperature $T_\mathrm{source}$. The power at the output of the receiver is calculated as 
\begin{equation} \label{eq:simple_power_gain}
    P_\mathrm{out} = k_\mathrm{B} B G_\mathrm{rec} (T_\mathrm{source} + T_\mathrm{rec}),
\end{equation}
where $B$ is the channel bandwidth, $k_\mathrm{B}$ is the Boltzmann constant, and the unknowns $G_\mathrm{rec}$ and $T_\mathrm{rec}$ denote the gain and noise temperature of the receiver, respectively.
The first-order correction is performed by Dicke switching \citep{Dicke_switching} between a cold load and a noise source, and measuring their power spectral densities (PSDs). The noise source and the cold load are assumed to be perfectly matched with an impedance of $50~\Omega$. Differencing the sources and arranging the terms would give us the source temperature $T^*_\mathrm{source}$ as 
\begin{equation} \label{Eq:1st_order_Dicke_switch}
    T^*_\mathrm{source} = T_\mathrm{NS} \left(\frac{P_{\mathrm{source}} - P_\mathrm{L}}{P_\mathrm{NS} - P_\mathrm{L}} \right) + T_\mathrm{L} ,
\end{equation}
where the PSDs $P_{\mathrm{source}}$, $P_\mathrm{L}$ and $P_\mathrm{NS}$ are measured for the source, (cold) load and noise source, respectively. The temperatures for load ($T_\mathrm{L}$) and noise source ($T_\mathrm{NS}$) are estimated later in the calibration process.

However, we need to include the effects of impedance mismatches in the system. 
Following \citet{Monsalve_2017_EDGES_receiver_calibration} and \citet{Bayesian_nwp_conjugate_priors}, the PSDs are propagated through the signal chain, including the effects of reflections, gains and losses in the path. 
These PSD terms are expanded in the Dicke switching equation to yield the calibrated source temperature $T_\mathrm{source}$.
We substitute $\theta_\mathrm{NS} \equiv T_\mathrm{NS}$ and $\theta_\mathrm{L} \equiv T_\mathrm{L}$, and Equation~\eqref{Eq:1st_order_Dicke_switch} expands as 
\begin{equation} \label{Eq:nwp_detailed_eq}
\begin{split}
    \theta_\mathrm{NS} \left(\frac{P_{\mathrm{source}} - P_\mathrm{L}}{P_\mathrm{NS} - P_\mathrm{L}} \right) + \theta_\mathrm{L} 
    &=
    T_{\mathrm{source}} \left[ \frac{1-\absval{\Gamma_{\mathrm{source}}}^2}{\absval{1- \Gamma_{\mathrm{source}} \Gamma_{\mathrm{rec}}}^2} \right] \\
    &+
    \theta_\mathrm{unc} \left[ \frac{\absval{\Gamma_{\mathrm{source}}}^2}{\absval{1- \Gamma_{\mathrm{source}} \Gamma_{\mathrm{rec}}}^2} \right] \\
    &+
    \theta_\mathrm{cos} \left[ \frac{\Re \left( \frac{\Gamma_{\mathrm{source}}}{1- \Gamma_{\mathrm{source}} \Gamma_{\mathrm{rec}}} \right)}{\sqrt{1- \absval{\Gamma_{\mathrm{rec}}}^2}} \right] \\
    &+
    \theta_\mathrm{sin} \left[ \frac{\Im \left( \frac{\Gamma_{\mathrm{source}}}{1- \Gamma_{\mathrm{source}} \Gamma_{\mathrm{rec}}} \right)}{\sqrt{1- \absval{\Gamma_{\mathrm{rec}}}^2}} \right]
    .
\end{split}
\end{equation}

Here $\theta_\mathrm{unc}$, $\theta_\mathrm{cos}$ and $\theta_\mathrm{sin}$ are defined as the noise-wave parameters.
The terms $\theta_\mathrm{NS}$ and $\theta_\mathrm{L}$ are included as unknown parameters to absorb the impedance mismatches at the receiver-noise source and receiver-cold load reference planes, respectively. 
All 5 terms are fitted in the calibration process \citep{Bayesian_nwp_conjugate_priors}. This can be simplified to give us the calibration equation
\begin{equation} \label{Eq:nwp_summing_equation}
    T_{\mathrm{source}} = X_\mathrm{unc} \theta_\mathrm{unc} + X_\mathrm{cos} \theta_\mathrm{cos} + X_\mathrm{sin} \theta_\mathrm{sin} + X_\mathrm{NS} \theta_\mathrm{NS} + X_\mathrm{L} \theta_\mathrm{L}, 
\end{equation}
which can be transformed to a shorthand matrix notation, while including instrument noise $\sigma$ for completeness as
\begin{equation} \label{Eq:nwp_matrix_eq-ultra-reduced_eq}
    T_\mathrm{source}(\nu) = \mathbf{X_\mathrm{source}}(\nu) \mathbf{\Theta}(\nu) + \sigma(\nu) .
\end{equation}
Here we note that all of the terms are a function of frequency $(\nu)$, the notation for which shall be omitted further in the paper for clarity. 
The matrices stand for 
\begin{equation*}
\begin{split}
    \mathbf{X}_\mathrm{source} &\equiv \begin{pmatrix}
        X_\mathrm{unc} & X_\mathrm{cos} & X_\mathrm{sin} & X_\mathrm{NS} & X_\mathrm{L}
    \end{pmatrix},
    \\
    \mathbf{\Theta} &\equiv \begin{pmatrix}
        \theta_\mathrm{unc} & \theta_\mathrm{cos} & \theta_\mathrm{sin} & \theta_\mathrm{NS} & \theta_\mathrm{L} 
    \end{pmatrix}^T    ,
\end{split}
\end{equation*}
where the X-terms for each source are constructed using the measured PSDs and reflection coefficient as 
\begin{equation}\label{Eq:X-matrix_def}
\begin{split}
    X_\mathrm{unc} =& - \frac{\absval{\Gamma_{\mathrm{source}}}^2}{1 - \absval{\Gamma_{\mathrm{source}}}^2}     ,
    \\
    X_\mathrm{cos} =& - \frac{\Re \left( \frac{\Gamma_{\mathrm{source}}}{1- \Gamma_{\mathrm{source}} \Gamma_{\mathrm{rec}}} \right) \absval{1- \Gamma_{\mathrm{source}} \Gamma_{\mathrm{rec}}}^2}
    {\left( 1 - \absval{\Gamma_{\mathrm{source}}}^2 \right) \sqrt{1- \absval{\Gamma_{\mathrm{rec}}}^2}}     ,
    \\
    X_\mathrm{sin} =& - \frac{\Im \left( \frac{\Gamma_{\mathrm{source}}}{1- \Gamma_{\mathrm{source}} \Gamma_{\mathrm{rec}}} \right) \absval{1- \Gamma_{\mathrm{source}} \Gamma_{\mathrm{rec}}}^2}
    {\left( 1 - \absval{\Gamma_{\mathrm{source}}}^2 \right) \sqrt{1- \absval{\Gamma_{\mathrm{rec}}}^2}}     ,
    \\
    X_\mathrm{NS} =& \left( \frac{P_{\mathrm{source}} - P_\mathrm{L}}{P_\mathrm{NS} - P_\mathrm{L}} \right) \frac{\absval{1- \Gamma_{\mathrm{source}} \Gamma_{\mathrm{rec}}}^2}{1 - \absval{\Gamma_{\mathrm{source}}}^2}     ,
    \\
    X_\mathrm{L} =& \frac{\absval{1- \Gamma_{\mathrm{source}} \Gamma_{\mathrm{rec}}}^2}{1 - \absval{\Gamma_{\mathrm{source}}}^2} .
\end{split}
\end{equation}

For the least-squares approach, we arrive at a system of linear equations using all calibrators, which can be equivalently represented as a matrix where each row corresponds to a calibrator.
The modified matrix, then, is 
\begin{equation} \label{eq:all_cal_calibration_eq}
    \mathbf{T} = \mathbf{X} \mathbf{\Theta} + \boldsymbol{\sigma} .
\end{equation}
The design matrix $\mathbf{X}$ at each frequency then has a shape of ($N_\mathrm{cal} \times 5 $), where $N_\mathrm{cal}$ is the number of sources in the calibrator set.
The problem of calibration now becomes solving this matrix and estimating the five unknown parameters $(\theta_\mathrm{unc}, \theta_\mathrm{cos}, \theta_\mathrm{sin}, \theta_\mathrm{NS}, \theta_\mathrm{L})$ 
at each frequency.

\subsection{The calibration sources}
\begin{figure}
    \centering
    \includegraphics[width=0.9\linewidth]{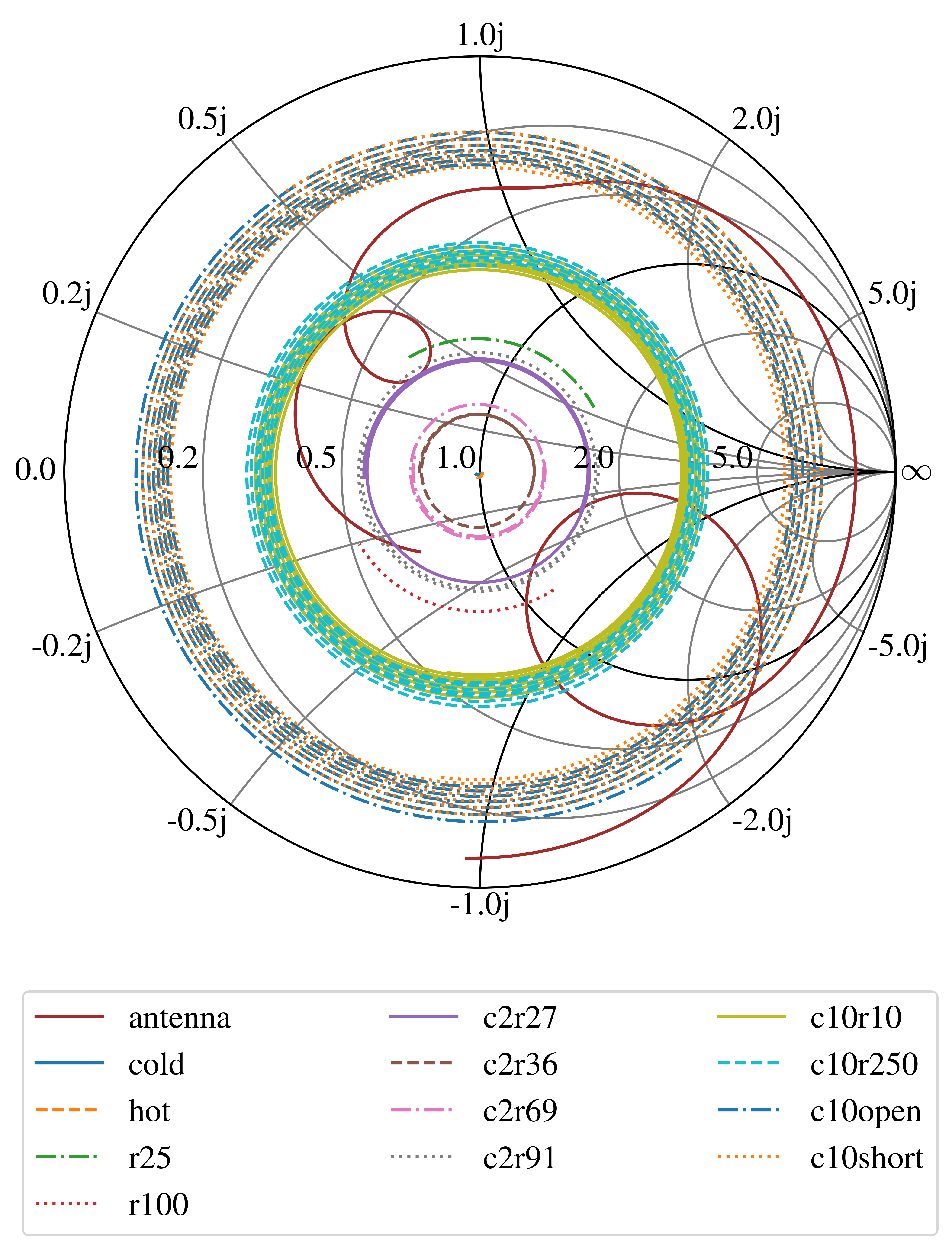}
    \caption{A Smith chart showing the reflection coefficients of the 12 calibrators used in REACH. The cold and hot loads are well matched, and their reflection coefficients are close to the centre of the plot. For reference, the antenna reflection coefficient is also plotted. The measurements are taken in the range of 50--130~MHz.} 
    \label{fig:Smith_chart_calibrators}
\end{figure}

The calibration equation in Equation~\eqref{Eq:nwp_summing_equation} with the five unknown noise-wave parameters can be solved by using calibration sources with known temperatures ($T_\mathrm{source}$) and reflection coefficients ($\Gamma_\mathrm{source}$).
REACH uses internal calibration sources, which are constructed by connecting a resistor on a cable. The calibrators are chosen to vary in temperature and complex reflection coefficient.
This is achieved by heating a $50~\Omega$ load, referred to as the hot load, and by varying cable lengths, providing good coverage over the Smith chart, as shown in Figure~\ref{fig:Smith_chart_calibrators}. 
The hot load sets an absolute temperature scale for all measurements, aiding in absolute calibration of the receiver, while the complex reflection coefficients are necessary for determining the spectral properties of the antenna.
An internal VNA in used to measure the reflection coefficients of the sources and the temperatures are measured using temperature probes, both \textit{in situ}. We further model this temperature to the reference plane at the input of the receiver, accounting for the s-parameters of the signal path \citep{REACH_radiometer_design}. 

The final calibrator set is as follows:
\begin{itemize}[wide, labelwidth=!,itemindent=!]
    \item a $50~\Omega$ cold load at ambient temperature 
    \item a $50~\Omega$ hot load at 370~K
    \item ambient $25~\Omega$ and $100~\Omega$ loads
    \item ambient $27~\Omega$, $36~\Omega$, $69~\Omega$ and $91~\Omega$ loads connected to a 2~m cable
    \item ambient Open and Short terminations and $10~\Omega$ and $250~\Omega$ loads connected to a 10~m cable
\end{itemize}

In this work, we use a shorthand notation for these sources where we have \texttt{hot} and \texttt{cold}, non-cabled sources denoted by their resistance - \texttt{r25} and \texttt{r100} - and the cabled source names preceded by their cable length - \texttt{c2r27}, \texttt{c2r36}, \texttt{c2r69}, \texttt{c2r91}, \texttt{c10r10}, \texttt{c10r250}, \texttt{c10open}, and \texttt{c10short}.

\subsection{Noise estimation} 
\subsubsection{Noise-dominated noise-wave parameter} \label{Sec:Noisy_theta_noise_estimation}
We use two methods of noise estimation in this paper. The first one assumes a similar PSD noise for all sources and noiseless s-parameter measurements of the path and sources. This is used to derive a relation between calibrator integration time and noise in the estimated noise-wave parameters, and thus in the final antenna solution.
 
Following the derivation in Appendix~\ref{apdx_noise_dominated_nwp}, for antenna integration time $\tau_\mathrm{ant}$, we arrive at the relation for the variance in the calibrated antenna temperature as
\begin{equation} \label{eq:final_Tant_noise}
    \sigma_\mathrm{ant}^2 = \mathbf{X}_\mathrm{{ant}} \mathbf{\Sigma_\Theta} \mathbf{X}_\mathrm{ant}^T 
    + \overline{\theta_\mathrm{NS}}^2 \frac{D}{\tau_\mathrm{ant}} ,
\end{equation}
where $\mathbf{X}_\mathrm{ant}$ corresponds to the design matrix for the antenna (or validation) source, $D$ is a constant and $\mathbf{\Sigma_\Theta}$ is the covariance matrix for the estimated noise-wave parameters. 
The overbar corresponds to the central noiseless value of the quantity.
The first term represents the uncertainty contributed by the estimated noise-wave parameters due to the noise in calibrator measurements, while the second term comes from the noise in antenna PSD measurements. 
The covariance matrix $\mathbf{\Sigma_\Theta}$ is given by
\begin{equation}\label{eq:Sigma_Theta}
    \mathbf{\Sigma_\Theta} 
    \approx 
    \frac{\overline{X_\mathrm{NS}}^2 b d}{\tau_\mathrm{cal}}
    \frac{\theta_\mathrm{NS}^2}{
    \left(\sum_{i=1}^{N_\mathrm{cal}}
    \overline{\mathbf{X}_i}^{T}\overline{\mathbf{X}_i}\right)} ,
\end{equation}
where $b$ and $d$ are constants and $\tau_\mathrm{cal}$ gives the integration time for each calibrator.

To facilitate comparison of a change in total calibration time for an optimised calibrator set, we define a reference condition, where the summation in the denominator scales with the number of calibrators. This modifies the denominator as
\begin{equation*}
\tau_\mathrm{cal} 
\left(
\sum_{i=1}^{N_\mathrm{cal}} 
\overline{\mathbf{X}_i}^{T} \overline{\mathbf{X}_i} 
\right) 
\approx 
\tau_\mathrm{cal} N_\mathrm{cal} \mathcal{M}, 
\end{equation*}
where $\overline{\mathbf{X}_i}^{T} \overline{\mathbf{X}_i} \approx \mathcal{M}, \forall i= 1,2, ... , N_\mathrm{cal}$, where the condition implies $\overline{\mathbf{X}_i}^{T} \overline{\mathbf{X}_i}$ is similar for all calibrators.
We note that this is only a reference condition and is not perfectly true due to impedance mismatches. However, the effect of variation evens out as we use calibrator sets with sources on opposite sides of the Smith chart.

This gives us the total calibration time $\tau_\mathrm{cal}^\mathrm{total} \equiv N_\mathrm{cal} \tau_\mathrm{cal}$.
We further assume the antenna noise term in Equation~\eqref{eq:final_Tant_noise} is negligible compared to the first term, giving us 
\begin{equation}
\begin{split}
    \sigma_\mathrm{ant}^2
    &\propto
    \frac{1}{N_\mathrm{cal} \tau_\mathrm{cal}}
    \\
    &\propto
    \frac{1}{\tau_\mathrm{cal}^\mathrm{total}}     .
\end{split}
\end{equation}

This gives a crude relation useful to assess the `goodness' of a calibration solution for a given total calibration time due to the respective calibrator set.

\subsubsection{Noise-less noise-wave parameter} \label{Sec:Minimum_noise_estimation}

The second method adds the assumption that the calibration provides noiseless noise-wave parameters $\Theta$ and includes possible cross-terms in the PSD covariance across Dicke switching sources. In this case, the first term in Equation~\eqref{eq:final_Tant_noise} vanishes and provides us with the minimum possible noise for a calibrated source temperature.
The equation is then derived by propagating noise through Equation~\eqref{Eq:nwp_detailed_eq} \citep{kirkham2025accountingnoisesingularitiesbayesian}, providing us the formula
\begin{equation} \label{Eq:Noise_propagation_and_estimation}
\begin{split}
    \left(\sigma^\mathrm{min}_{T_\mathrm{source}}\right)^2  &=  \left( \frac{\theta^\mathrm{fit}_\mathrm{NS} X_\mathrm{L}}{P_\mathrm{NS} - P_\mathrm{L}} \right)^2 
    \Biggl[  \sigma^2_{P_\mathrm{source}} + \sigma^2_{P_\mathrm{L}} - 2 \sigma_{P_\mathrm{source} P_\mathrm{L}}  \\
    &+ \left( \frac{P_\mathrm{source} - P_\mathrm{L}}{P_\mathrm{NS} - P_\mathrm{L}} \right)^2 (\sigma^2_{P_\mathrm{L}} + \sigma^2_{P_\mathrm{NS}} - 2 \sigma_{P_\mathrm{L} P_\mathrm{NS}}) \\
    &- 2 \left( \frac{P_\mathrm{source} - P_\mathrm{L}}{P_\mathrm{NS} - P_\mathrm{L}} \right) \sigma_{AB} 
    \Biggr] .
\end{split}
\end{equation}
Here $\theta^\mathrm{fit}_\mathrm{NS}$ is a fit to the measured $\theta_\mathrm{NS}$, $A = P_\mathrm{source} - P_\mathrm{L}$ and $B = P_\mathrm{NS} - P_\mathrm{L}$ and $\sigma_{IJ}$ is the covariance of $I$ and $J$.

As this metric sets a minimum threshold for noise in a calibrated solution, it shall be used for identifying acceptable solutions later in this work.

\subsection{Condition numbers} \label{sec:condition_num_intro}
Following the Equations~\ref{Eq:nwp_matrix_eq-ultra-reduced_eq} and~\ref{eq:all_cal_calibration_eq}, the noise-wave parameters can be estimated using linear least-squares approximation, with the assumption that the instrument noise $\sigma$ is similar for all the calibrators. 
Thus, for a given set of calibrator observations, the method attempts to minimize the loss
$L(\mathbf{T}, \mathbf{\Theta}) = \norm{\mathbf{T}- \mathbf{X} \mathbf{\Theta}}^2 $.  
Setting the gradient of the loss to zero, we get the best estimate $\hat{\mathbf{\Theta}}$ as
\begin{equation} \label{Eq:LLS_estimator}
    \hat{\mathbf{\Theta}} = (\mathbf{X}^T \mathbf{X})^{-1} \mathbf{X}^T \mathbf{T} = \mathbf{X}^+ \mathbf{T}  ,
\end{equation}
where we use the definition of Moore-Penrose inverse (or pseudo-inverse) as $\mathbf{X}^+ \equiv (\mathbf{X}^T \mathbf{X})^{-1} \mathbf{X}^T$ \citep{Moore_pensore_inverse}. This solution is derived elaborately in Equation~\eqref{app_eq:noiseless_lsq_minimization}.

Given Equation~\eqref{Eq:LLS_estimator}, the condition number $\kappa(\mathbf{X})$ can be used to establish a relationship between the uncertainty of $\mathbf{T}$ and the maximum error in estimation of $\mathbf{\Theta}$, based on how accurately the pseudo-inverse $\mathbf{X}^+$ can be calculated \citep{Condition_number}. More formally, for error $\epsilon$ in the measured temperatures in $\mathbf{T}$, and error $\mathbf{X}^+ \epsilon$ in $\mathbf{\Theta}$, the condition number can be derived as
\begin{equation} \label{Eq:Derive_cn_basic}
\begin{split}
   \kappa(\mathbf{X}) &= 
   \max_{\epsilon,\mathbf{T} \neq 0} 
   \left\{ 
   \frac{\norm{\mathbf{X}^+ \epsilon}}{\norm{\mathbf{X}^+ \mathbf{T}}} 
   /
   \frac{\norm{\epsilon}}{\norm{\mathbf{T}} } 
   \right \} 
\\
   &=  \max_{\epsilon \neq 0} \left\{ \frac{\norm{\mathbf{X}^+ \epsilon}}{\norm{\epsilon}}\right \} 
   \max_{\mathbf{T} \neq 0} 
   \left\{
   \frac{\norm{\mathbf{T}}}{\norm{\mathbf{X}^+ \mathbf{T}} } 
   \right \} 
   \\
   &=  \max_{\epsilon \neq 0} \left\{ \frac{\norm{\mathbf{X}^+ \epsilon}}{\norm{\epsilon}}\right \} 
   \max_{\mathbf{\Theta} \neq 0} \left\{\frac{\norm{\mathbf{X} \mathbf{\Theta}}}{\norm{\mathbf{\Theta}} } \right \}  
   \\
    &= \norm{\mathbf{X}^+} \norm{\mathbf{X}} ,
\end{split}
\end{equation}
where we have $\lVert\cdot\rVert$ defined as an $L^2$ norm and is sub-multiplicative, such that $\norm{AB} \leq \norm{A} \norm{B}$.

For $\kappa$ equal to one, we imply that the estimation error in the solution $\mathbf{\Theta}$ is no worse than that in the measurement of $T_\mathrm{source}$. Thus, for an equation, a lower condition number is favourable, referring to it as well-conditioned, whereas a higher value indicates it is ill-conditioned and might lead to large numerical errors.

\begin{figure*}
    \centering
    \includegraphics[width=1\linewidth]{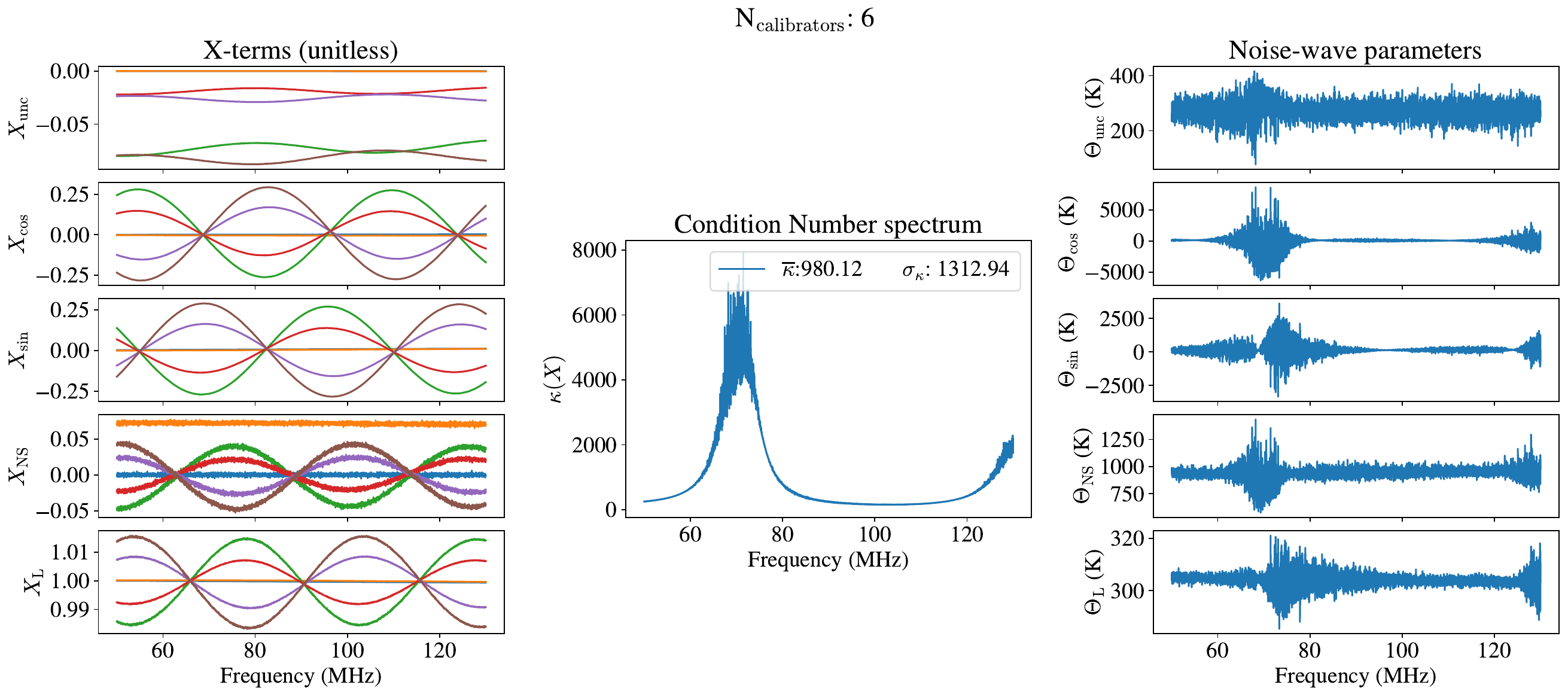} \\
    \vspace{0.5cm}
    \includegraphics[width=1\linewidth]{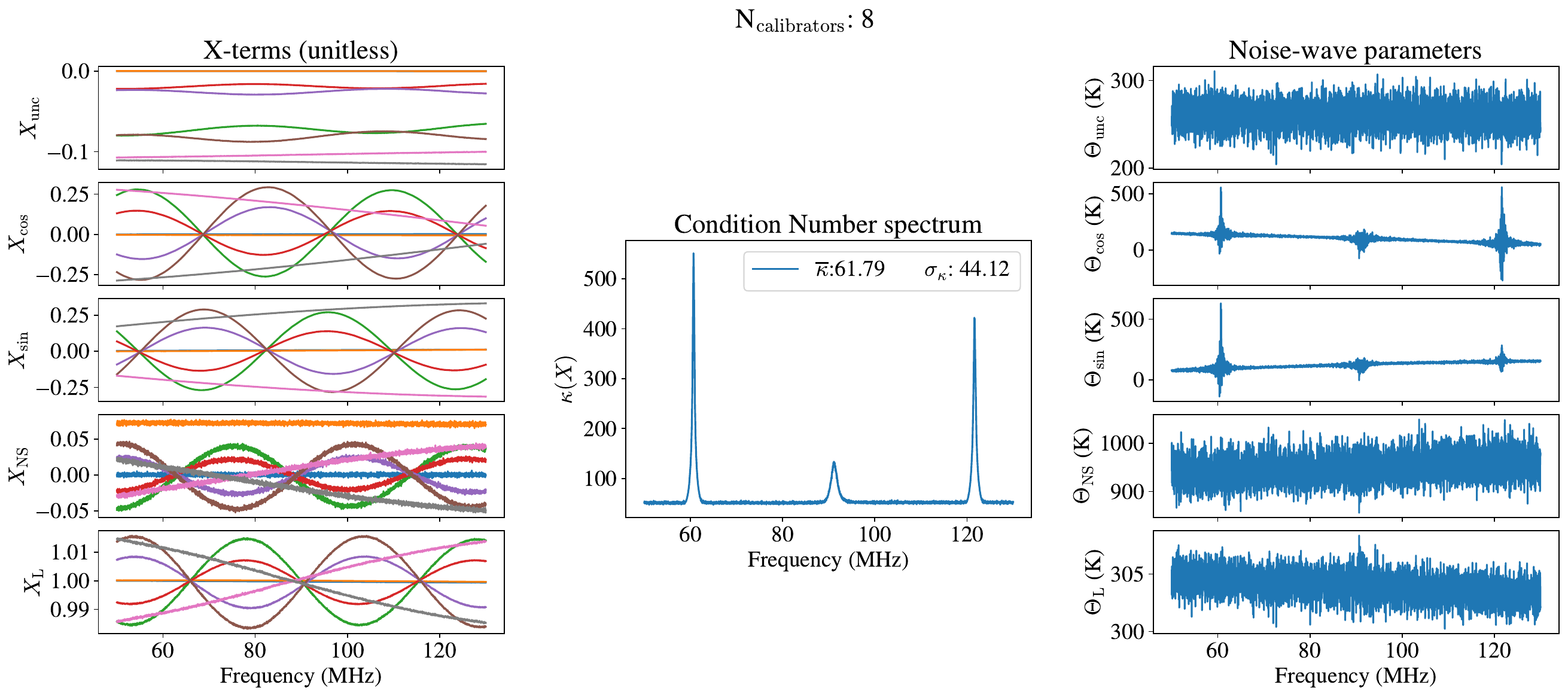}
    \caption{Plots for two cases of calibrator set sizes: \textit{(top)} 6 sources (\texttt{[cold, hot, c2r27, c2r36, c2r69, c2r91]}) and \textit{(bottom)} 8 sources (\texttt{[cold, hot, c2r27, c2r36, c2r69, c2r91, r25, r100]}). For each case, we show the X-terms \textit{(left)}, the respective condition number spectrum \textit{(centre)} and the estimated noise-wave parameters \textit{(right)}. In the X-terms plots for 8 calibration sources, the magenta and grey curves correspond to the additional calibrators in the set. It must be noted that the y-axis range is different for all noise-wave parameter plots.} 
    \label{fig:X-terms_and_cn_plot}
\end{figure*}

With the above definitions, it is possible to calculate $\kappa(\mathbf{X})$ for a given matrix $\mathbf{X}$, where $\mathbf{X}$ is a function of the calibrator set, constructed from the reflection coefficients $\Gamma_\mathrm{source}$ and the PSDs of the calibration sources, as well as the receiver reflection coefficient $\Gamma_\mathrm{rec}$, as in Equation~\eqref{Eq:X-matrix_def}. Two examples with different calibrator sets and the respective $\mathbf{X}$ are shown in Figure~\ref{fig:X-terms_and_cn_plot}, highlighting the effects of calibrator choice on the respective $\kappa(\mathbf{X})$ and the estimated noise-waves using the calibrator set.
 
We interpret Figure~\ref{fig:X-terms_and_cn_plot} using the concept of linear dependence. A given sequence of vectors $\mathbfit{v}_1, \mathbfit{v}_2 ..., \mathbfit{v}_n$ (or the columns or rows of a matrix) is said to be linearly dependent, if there exists scalars $a_1, a_2, ..., a_n$, not all zeros, which can be used to denote any vector as a linear combination of the other vectors in the set, as
\begin{equation*}
    \mathbfit{v}_k = \sum_{i=1, i \neq k}^n a_i \mathbfit{v}_i  .
\end{equation*}
Equivalently, a matrix constructed using these vectors will have a zero determinant, be non-invertible and have an infinite condition number. 
However, in computing, we are limited by the machine precision, thus leading to what we define as `near-linear dependence'.
Similarly, when vectors cannot be represented in this manner, they are said to be linearly independent, or `near-linearly independent', including the effects of machine precision.
Thus, in computing, the condition number in a way represents how `well-spread' the constituent vectors are in different directions, as used in \citet{Well_spread_points_SKA_low} and \citet{Well_spread_points_OSLC}.

For a given set of calibrators, the design matrix $\mathbf{X}$ is derived at each frequency using the relations in Equations~\ref{Eq:nwp_matrix_eq-ultra-reduced_eq}--\ref{eq:all_cal_calibration_eq}. The frequencies at which the columns (coefficient X) or rows (respective calibrator) become near-linear dependent (or become degenerate), the condition number shoots up. These frequencies are prone to large numerical error in the computation of the pseudo-inverse $\mathbf{X}^+$ or the inverse of the Gram matrix $(\mathbf{X}^T \mathbf{X})^{-1}$.
A matrix with `well-spread-out' elements would help overcome this near-linear dependence, which is attained by increasing the diversity of calibrators (and hence the X-terms) with different reflection coefficients and PSDs. 
This can be observed in the bottom panel of Figure~\ref{fig:X-terms_and_cn_plot} as the addition of two new sources in the X-matrix \textit{(left)} reduces the condition number \textit{(centre)}. This effect is further carried in the estimation of the noise-wave solutions, which can be observed in the right panel of the same figure. 

For further analysis and simplicity, we require a metric which can be used to assess the quality of a condition number spectrum.
A reasonable option would be a measure of central tendency $\overline{\kappa}$. This was chosen to be the mean of the spectrum due to its sensitivity to extreme values, such that high condition number values are strongly reflected in the metric.
Another metric considered was the standard deviation of the condition number spectrum $\sigma_\kappa$. 
Both of these quantities roughly encode the central value and smoothness of the condition number spectrum as they are weakly related to each other due to the inherent structure of the spectrum, as is evident from the $\kappa(\mathbf{X})$ plots in the centre panel of Figure~\ref{fig:X-terms_and_cn_plot}. Hence, minimizing one would be enough to get an overall smooth and low-valued condition number spectrum for a given calibration equation. The mean ($\overline{\kappa}$) was chosen for this purpose.

The aim is to establish a relationship between the `goodness' of calibration and $\overline{\kappa}$ of the calibration equation for different calibrator sets. This is detailed in the following sections.

\section{Application to REACH data} \label{sec:Application_to_reach_data}

In this section, we relate condition number of the design matrix for the receiver with the calibration solutions. 
We first study the trends between the mean condition number and calibration quality metrics.
This is followed by the calibrator set optimisation methods based on condition numbers.

\subsection{Condition number and quality of calibration}

To establish a relation between the condition number and calibration solutions, and simultaneously validate our tests and results, we require a validation source with a known temperature. For this purpose, we use one of the 12 sources in the receiver and calibrate the receiver using any combination of the remaining sources. Measurements and calibration are performed on the REACH radiometer. The methodology is as follows.

For a given source chosen as the validator, the remaining 11 sources were used to make a calibrator set of size varying from 5 to 11, giving us a total of $\sum_{i=5}^{11} {}^{11}C_i$ combinations. The lower limit of five is chosen, considering that the five unknown noise-wave parameters can be solved only with a minimum of five simultaneous linear equations. These sets were used to construct the matrix $\mathbf{X}$ and estimate $\mathbf{\Theta}$, following Equation~\eqref{eq:all_cal_calibration_eq}. The estimated $\mathbf{\Theta}$ is put in Equation~\eqref{Eq:nwp_matrix_eq-ultra-reduced_eq} to solve for the temperature $T_\mathrm{solution}$ of the validation source. 
This can be compared with the measured temperature of the validation source, which is referred to as the target temperature $T_\mathrm{target}$.
For each validation source and calibrator combination, we note the standard deviation $\sigma_\mathrm{T}$ of the temperature solution and the absolute mean deviation $\absval{\overline{\Delta T}}$ of the solution from the target, where $\Delta T (\nu) = T_\mathrm{solution}(\nu) - T_\mathrm{target}(\nu)$.
The standard deviation here encompasses both random noise and systematic error, in the form of unintended spectral features, in the calibrated temperature solution.
The standard deviation and absolute mean deviation are considered as measures of bandpass (spectral) and absolute calibration, respectively. 

\begin{figure*}
    \centering
    \includegraphics[width=0.99\linewidth]{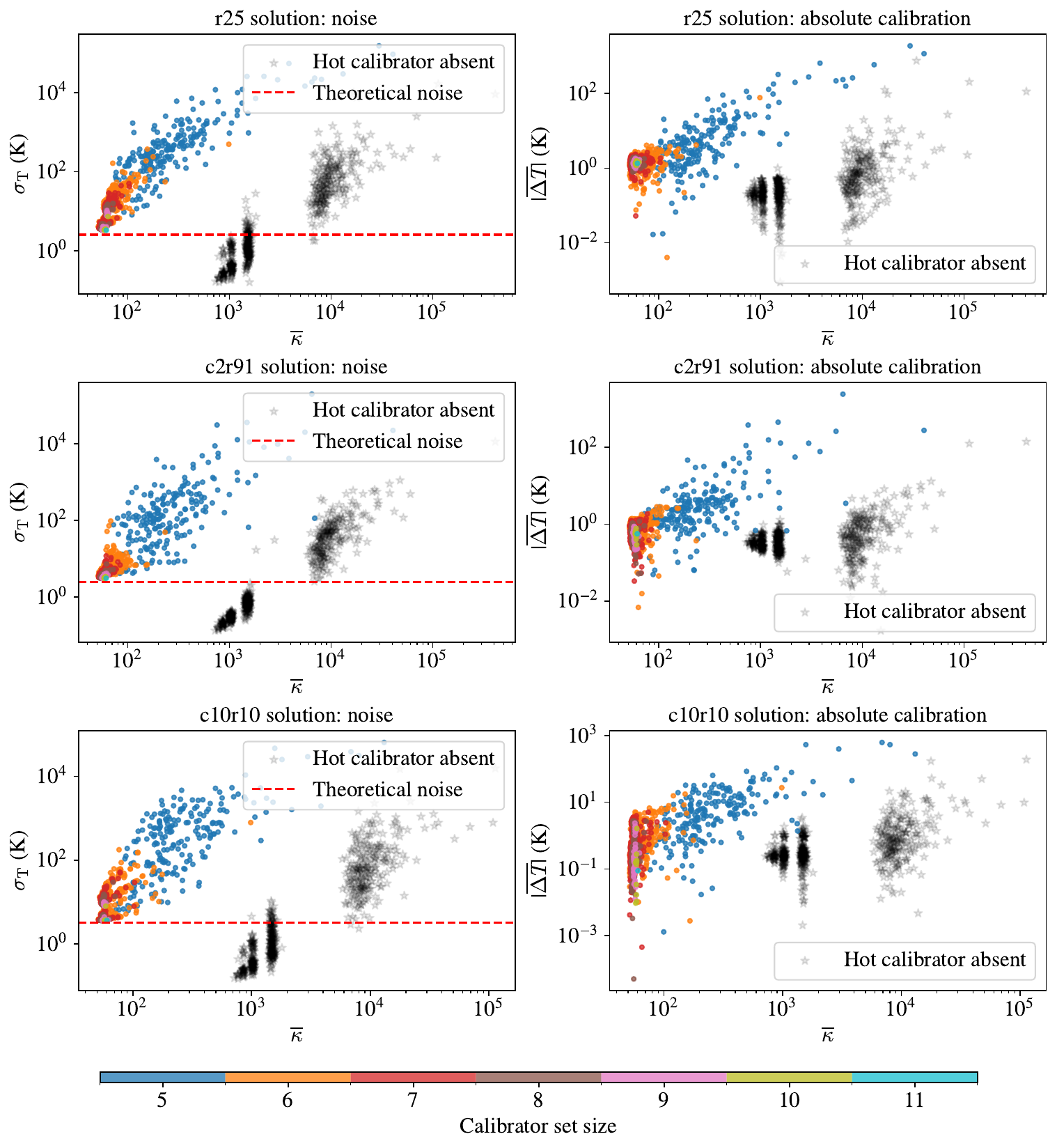}
    \caption{Calibration quality vs mean condition number plots for three validation sources with different cable lengths \textit{(top, middle and bottom rows respectively)}.
    For each validator, on the left we show the trends of standard deviation $\sigma_\mathrm{T}$ of the temperature solution and on the right, the absolute mean deviation $\absval{\overline{\Delta T}}$ of the solution from the target temperature.
    Each point in the figure represents a different calibrator set with sizes as shown in the colour bar at the bottom. The black $\star$ shows the calibrator sets which do not include the hot source. 
    An estimate of (minimum) theoretical noise, calculated using an 11-calibrator set in Equation~\eqref{Eq:Noise_propagation_and_estimation}, is also shown as a red dashed line in the plots on the left column for each validation source.} 
    \label{fig:T_source_std_dev_and_cn_relation}
\end{figure*}

In Figure~\ref{fig:T_source_std_dev_and_cn_relation}, we plot the variation of the standard deviation $\sigma_\mathrm{T}$ and absolute mean deviation $\absval{\overline{\Delta T}}$ of the solution with the mean condition number $\overline{\kappa}$ for three validation sources. Each point in the plot represents a different calibrator set with sizes varying between 5 and 11, specified by the colour code.
The plot also shows the expected minimum noise $\sigma^\mathrm{min}_{T_\mathrm{source}}$ in the solution for each validation source as explained in Section~\ref{Sec:Minimum_noise_estimation}. For the respective validator, $\theta_\mathrm{NS}^\mathrm{fit}$ in Equation~\eqref{Eq:Noise_propagation_and_estimation} is estimated using all 11 calibrators .

We make a few observations from Figure~\ref{fig:T_source_std_dev_and_cn_relation}.
Firstly, $\sigma_\mathrm{T}$ tends to increase with $\overline{\kappa}$ for the calibrator set. However, the points span out at higher $\overline{\kappa}$, which are usually associated with smaller calibrator sets. These are a result of spectral structures in the condition number array and thus the solution, which are not always captured perfectly by the mean or standard deviation.
Second, we note the multiple groups of points in the plot. A deeper investigation revealed that the group with the lowest mean condition number (the leftmost group) consistently included the hot calibration source, while others mostly did not (marked in black). A significant number of calibrator combinations missing \texttt{hot} also gave solutions with noise less than the minimum theoretical estimate. 
The effect of a missing hot calibrator was usually observed as poorly estimated $\theta_\mathrm{NS}$ and a strong spectral distortion or bad absolute calibration in $T_\mathrm{source}$.
\textit{These results set the hot source as a necessary calibrator, especially for absolute calibration.}


While a similar increasing trend is not observed in the case of absolute calibration, it was observed that the calibrator sets of sizes 6-8 with low $\overline{\kappa}$ provided lower $\absval{\overline{\Delta T}}$ compared to those with higher $\overline{\kappa}$ values. 
Additionally, the absolute calibration strongly depends on the measurement accuracy of the source temperatures and the temperature of the receiver.
Other sources of error include modelling of source temperatures and cable parameters, errors in PSD and s-parameter measurements, as well as drift in the instrument.
We note that all of these are beyond the scope of condition number formalism.
Absolute mean deviation, however, will be noted in the final calibration results.


The above trends were observed across multiple datasets and different validation sources. This increased confidence in treating the mean condition number $\overline{\kappa}$ as a useful metric for estimating the goodness of calibration, provided the calibrator set includes the hot source.

\subsection{Selection of preferred calibrators} \label{sec:best_set}
As a lower mean condition number is preferred for better calibration, all possible calibrator sets can be sorted based on respective $\overline{\kappa}$ to get the best sets. 
Once we have ordered all calibrator sets based on increasing $\overline{\kappa}$ values, we can perform a statistical analysis to identify the favourite calibrators for the REACH receiver.
For each validator, all the respective calibrator sets yield $\overline{\kappa}$ values starting from around 40--50. This value increases non-uniformly as we go over the calibrator sets. To set a good contrast in calibrator occurrence and include only good calibrator sets, we set an upper limit of 53 for $\overline{\kappa}$ value based on visual inspection of the smoothness of the solution spectra.
The `favourite' calibrators for each respective validation source can then be found by counting the number of times each calibration source appears in these (good) sets.

\begin{figure}
    \centering
    \includegraphics[width=0.9\linewidth]{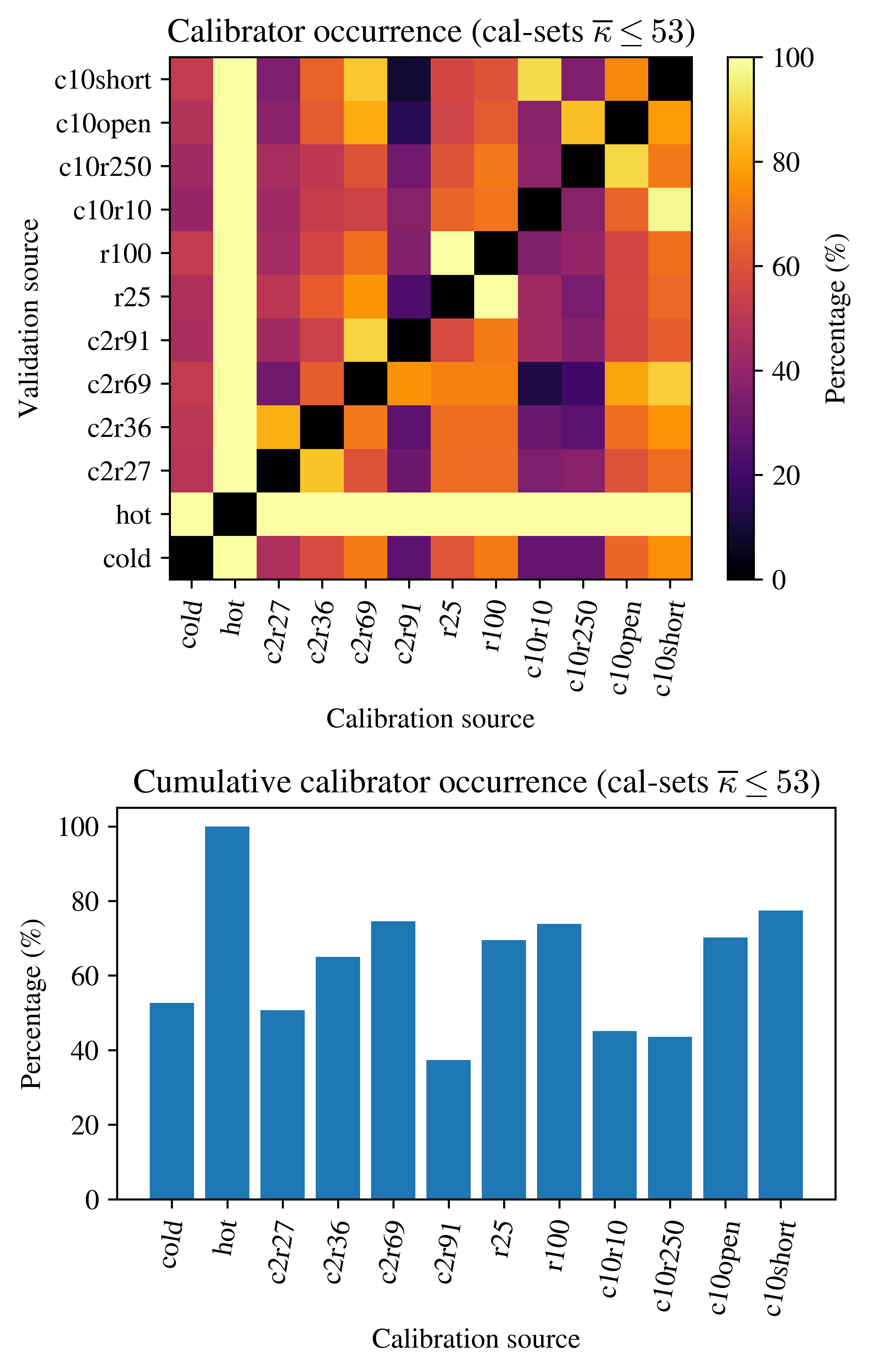}
    \caption{\textit{Top:} For each validation source (y-axis), the intensity chart shows the percentage occurrence of each calibrator in the best calibrator sets with $\overline{\kappa}~\leq~53$. \textit{Bottom:} A cumulative percentage occurrence of each calibrator obtained by averaging the respective column in the top chart.}
    \label{fig:best_cals_percentage_occurence} 
\end{figure}

We present the percentage occurrence of these calibrators in the top panel of Figure~\ref{fig:best_cals_percentage_occurence}. For a given validation source, the calibrator set combination was created from the pool of sources without the validation source; hence, a zero occurrence is observed across the diagonal of the grid plot. 
The hot source was present in all the sets, whereas sources like \texttt{c2r91}, \texttt{c10r10}, and \texttt{c10r250} were the least used sources. The figure also shows the cumulative percentage occurrence for the number of times each calibration source was used across all validators, representing similar information.

This method was extended by including all 12 sources for generating combinations of calibrator sets, in which case, the sets would be used to calibrate an antenna source (or an external source at the input of the receiver) as the target source.
It was observed that the calibrator set \texttt{[hot, r100, c2r36, c2r69, c10open, c10short]} yielded the lowest mean condition number $\overline{\kappa}$, which was consistent for multiple datasets.
As expected from the discussion in Section~\ref{sec:condition_num_intro}, the best calibrator set consists of sources that provide a diversity in PSD and impedance (through temperature, termination resistance and cable lengths), while maintaining near-linear independence.
%
%
\textit{This set shall henceforth be referred to as the `optimised set' and chosen for calibration of the receiver.} The respective results are presented later in Section~\ref{sec:calibration_results}. 

\begin{figure}
    \centering
    \includegraphics[width=0.9\linewidth]{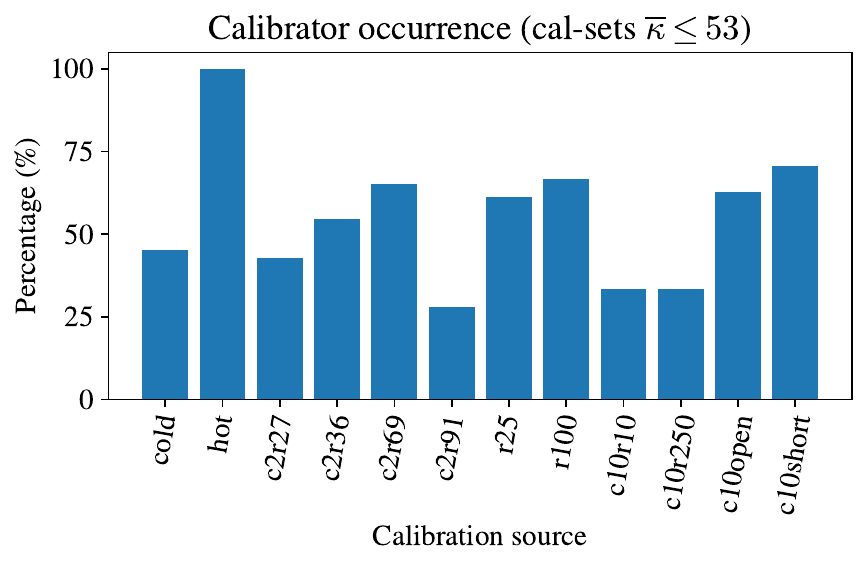}
    \caption{Distribution of calibrator occurrence percentage in the best sets yielding the lowest $\overline{\kappa}$ with the constraint $\overline{\kappa}~\leq~53$. All 12 available sources were used to make combinations for the calibrator sets.}
    \label{fig:Best_ant-cals}
\end{figure}

Similar to the previous case, a statistical analysis can be performed to get the favourite or least favourite calibrators for antenna calibration.
In Figure~\ref{fig:Best_ant-cals}, we present the percentage occurrence of the calibrators in the calibrator sets with the lowest $\overline{\kappa}$ with the constraint $\overline{\kappa}~\leq~53$.
The source \texttt{c2r91} was identified as one of the least preferred calibration sources. This was therefore chosen to be the validation source for most of the analysis. 

\subsection{Piecewise calibration} \label{sec:piecewise_intro}
\begin{figure}
    \centering
    \includegraphics[width=0.9\linewidth]{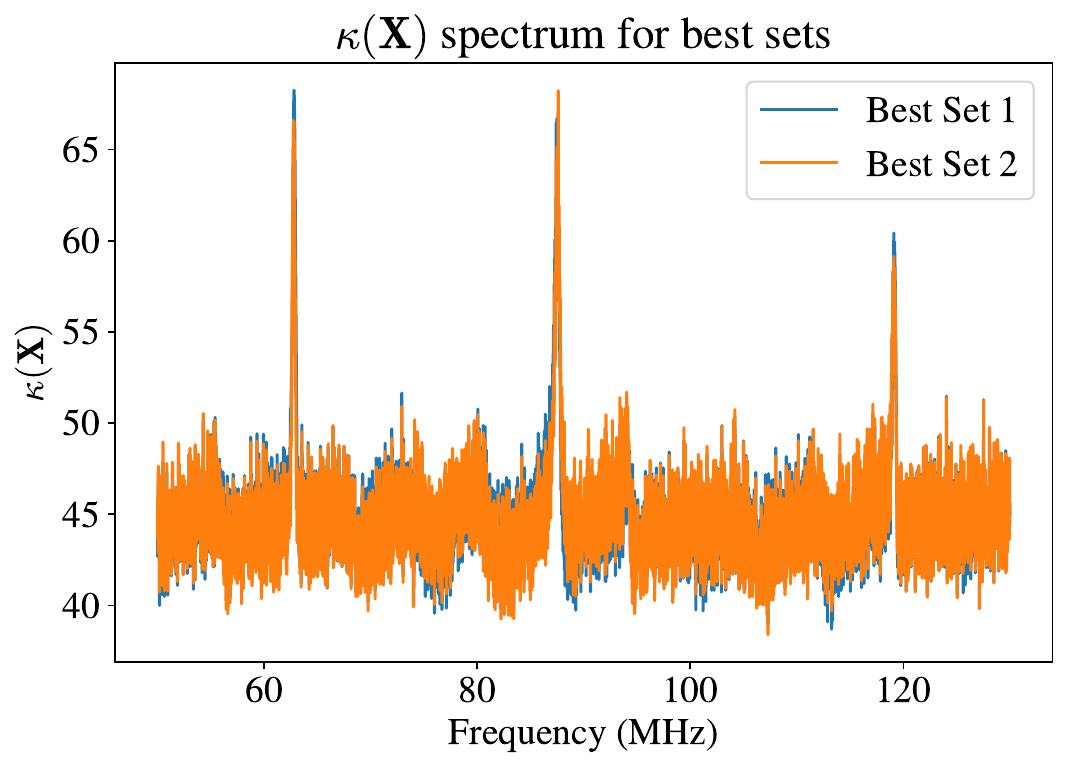}
    \includegraphics[width=0.9\linewidth]{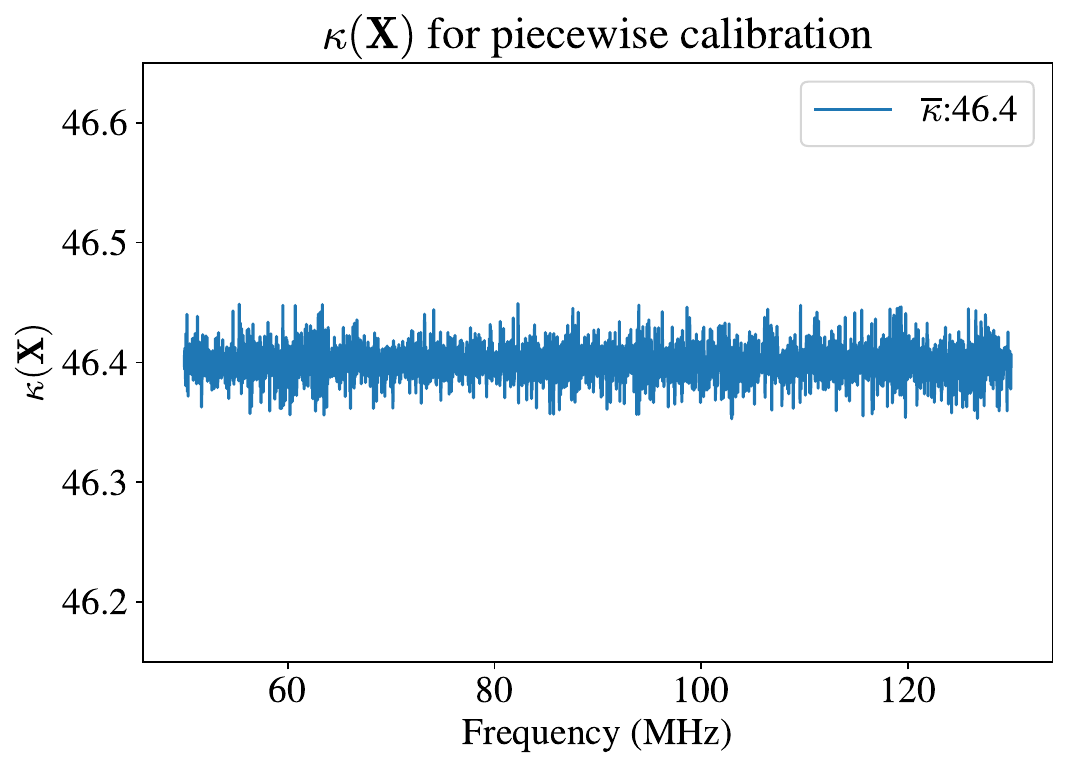}
    \caption{\textit{(Top:)} The condition number spectrum for the best two calibrator sets providing the lowest $\overline{\kappa}$ for a certain dataset, when the complete pool of calibrators was chosen for making combinations. The calibrator set used were \texttt{[hot, c2r36, c2r69, r100, c10open, c10short]} and \texttt{[hot, c2r36, c2r69, r100, c10r250, c10short]}, for Set 1 and 2  respectively. \textit{(Bottom:)} For the same dataset, the condition number spectrum after enforcing smoothness on the spectrum. The smoothness parameters chosen were $\overline{\kappa} = 46.4$ and $\kappa_\mathrm{win} = 0.05$. Out of 6553 frequency channels, 84 were flagged.} 
    \label{fig:cn_ripples}
\end{figure}

While the mean of the condition number served as a good central tendency to approximate a high or low condition number for the full REACH band of 50--130~MHz, it was observed that its spectrum could contain narrow peaks, as well as low-lying ripples, as visible in the example plots in the top panel of Figure~\ref{fig:cn_ripples}. 
This could degrade the calibration quality at those frequency points.
A way to overcome these high condition number peaks was by dividing the entire frequency band into smaller pieces and getting the best calibrator set for each sub-band, while enforcing a smoothness constraint over the condition number in each band. 
Any sub-band which would not be able to provide such a calibrator set would be flagged for the latter pipeline.

To avoid excessive flagging, we chose the size of each sub-band to be one frequency channel ($\sim~$12.208~kHz for REACH).
The smoothness was then enforced by providing a mean value $\overline{\kappa}$ and a threshold deviation $\kappa_\mathrm{win}$ from the mean. 
The value of these parameters was optimised by observing the calibration solution of the validator (here \texttt{c2r91}) for each frequency channel, while varying the calibrator set.
The optimisation aimed to minimize the previously defined calibration metrics, $\sigma_\mathrm{T}$ and $\absval{\overline{\Delta T}}$, while keeping the number of channels flagged minimal, during the process.

An example of this implementation is shown in the bottom panel of Figure~\ref{fig:cn_ripples}, where the smoothness constraint parameters are $\overline{\kappa} = 46.4$ and $\kappa_\mathrm{win} = 0.05$.
The calibrated solution spectrum was observed to be sensitive to these parameters in the following way. The mean (central) value $\overline{\kappa}$ determines the number of channels are included within the threshold window. 
Similarly, $\kappa_\mathrm{win}$ imposes the smoothening constraint on condition number spectrum, at the expense of increased flagging of frequency channels.

We note that this method is a dynamic process, and the choice of calibrators and the optimised values of the smoothness constraint parameters shall vary on the dataset and the available pool of sources to create the calibrator sets.
Additionally, as each channel (sub-band) is calibrated with a different calibrator set with varying sizes, the noise-wave parameters obtained using least-squares are expected to exhibit different noise levels in each channel, as indicated by Equation~\eqref{eq:Sigma_Theta}.

We shall present the calibration result for this method later in Section~\ref{sec:calibration_results}, demonstrating its efficacy.

\section{Calibration results} \label{sec:calibration_results}
\begin{figure}
    \centering
    \includegraphics[width=1\linewidth]{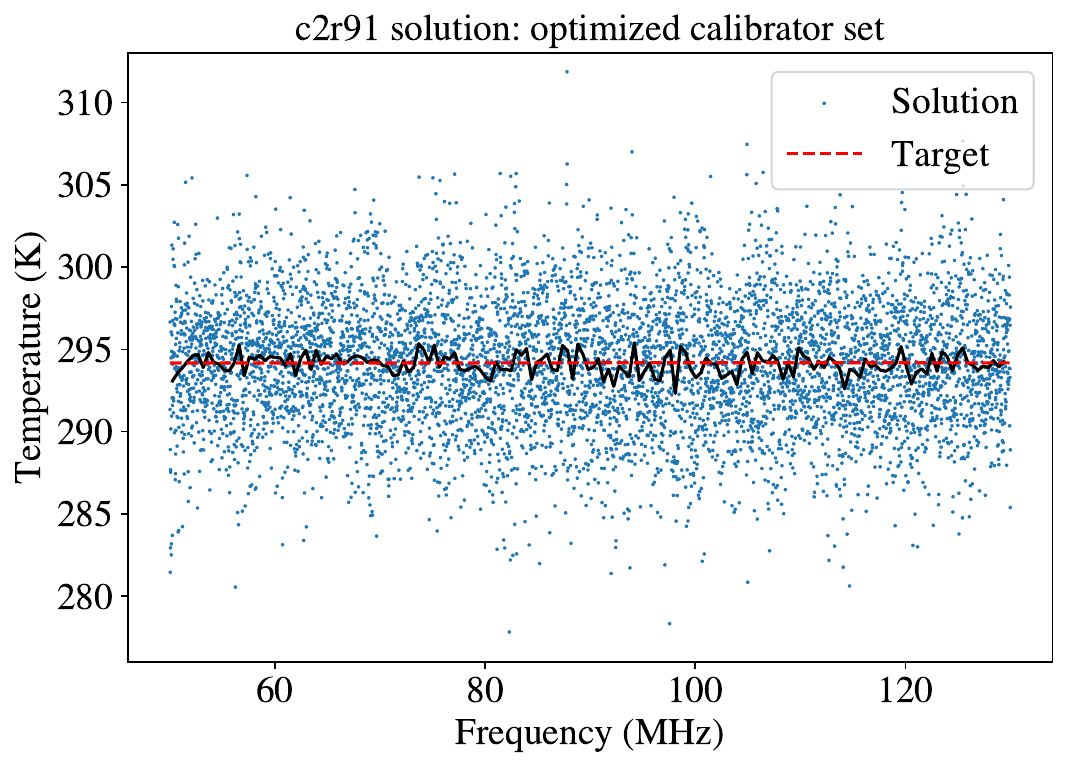}
    \caption{Calibration solution for the validation source \texttt{c2r91} using the optimised calibrator set \texttt{[hot, r100, c2r36, c2r69, c10open, c10short]}. The blue points represent the solution with a frequency resolution of 12.208~kHz. The black line is the binned solution with bin-width $\sim$~500~kHz. The red dashed line is the measured temperature using a probe on the resistor.}
    \label{fig:opt_cal_solution}
\end{figure}

We present calibration results of a measurement done on the REACH radiometer, deployed in the Karoo Desert. 
Figure~\ref{fig:opt_cal_solution} shows the calibrated temperature solution of the validation source \texttt{c2r91}, calibrated using the optimised set with the least mean condition number, as mentioned in Section~\ref{sec:best_set}.
In Table~\ref{table:compare_cal_sols}, the `goodness' of calibration is presented in the form of standard deviation $\sigma_\mathrm{T}$ and absolute mean deviation $\absval{\overline{\Delta T}}$ for calibrated \texttt{c2r91} temperature using three different calibration methods as follows:
\begin{enumerate}[label=(\Alph*), wide, labelwidth=!,itemindent=!]
    \item The remaining 11-calibrator set: \texttt{[cold, hot, r25, r100, c2r27, c2r36, c2r69, c10r10, c10r250, c10open, c10short]},
    \item The optimised 6-calibrator set: \texttt{[hot, r100, c2r36, c2r69, c10open, c10short]}, and
    \item Piecewise calibration, introduced in Section~\ref{sec:piecewise_intro}, where each frequency channel has a different optimised set.
\end{enumerate}
As discussed in Appendix~\ref{apdx_noise_dominated_nwp}, the noise in the calibrated source temperature varies inversely with the per-calibrator integration time $\tau_\mathrm{cal}$ and the number of calibrators $N_\mathrm{cal}$. 
Thus, to make a good comparison, based on the assumptions for the reference condition, introduced in Section~\ref{Sec:Noisy_theta_noise_estimation}, in the last column of the Table~\ref{table:compare_cal_sols}, we calculate normalised noise, had the total calibration time $\tau_\mathrm{cal}^\mathrm{total}$ been the same for all three cases.
For Case C, this becomes a frequency-by-frequency scaling as each channel is calibrated using a different calibrator set with variable sizes, and hence the total calibration time.

\begin{table*}
\centering
\caption{The calibration cases and the respective `goodness' of calibration are shown for a dataset measured on the REACH instrument with a frequency resolution of 12.208~kHz. 
The total calibration time is calculated based on the integration time of the calibration source (i.e. 30~sec~$\times~N_\mathrm{cal}$) and the noise source and reference load observation for Dicke switching is ignored for simplicity, as it would only multiply a factor of 3 to all $\tau_\mathrm{cal}^\mathrm{total}$ values.
The piecewise optimised set is chosen based on the discussion in Section~\ref{sec:piecewise_intro}, and 84 of 6553 channels are flagged in the calibrated spectrum.
The last column shows the normalised noise for $\tau_\mathrm{cal}^\mathrm{total}~=~300$~sec, where the normalisation is done based on the assumption of a reference condition, introduced in Section~\ref{Sec:Noisy_theta_noise_estimation}.}
\begin{tabular}{| m{0.03\linewidth} | m{0.18\linewidth} | m{0.15\linewidth} | m{0.07\linewidth} | m{0.15\linewidth} | m{0.15\linewidth} |} 
\hline
 \textbf{Case} & \textbf{Calibration set} & \textbf{Total calibration time $\tau_\mathrm{cal}^\mathrm{total}$} & \textbf{Noise $\sigma_\mathrm{T}$} & \textbf{Absolute calibration $\absval{\overline{\Delta T}}$}  & \textbf{$\sigma_\mathrm{T}$ normalised for $\tau_\mathrm{cal}^\mathrm{total}~=~300~$sec}\\ 
 \hline
 A & 11 sources & 330 sec & 3.248~K & 0.214~K & 3.407~K \\  
 \hline
 B & Optimised set (6 sources) & 180 sec & 3.755~K & 0.076~K & 2.908~K\\
 \hline
 C & Piecewise optimised set & Variable & 4.016~K & 0.005~K & 3.249~K \\
 \hline
\end{tabular}
\label{table:compare_cal_sols}
\end{table*}
    
For normalised noise, it was observed that both the full-band optimised calibrator set and piecewise optimised calibration performed better than the usual 11-calibrator case, with a reduced noise of approximately 15~\% and 5~\%, respectively, for a similar total calibration time. This was found to be consistent with several datasets observed on REACH.
Moreover, we observed that the spectral structure in the binned spectrum remained similar for all three methods, suggesting that the new methods did not introduce any additional dominant systematic to the calibration.
Condition number-based optimisation also yielded better absolute calibration for this measurement by an order of magnitude. This, however, was observed to fluctuate over several datasets within approximately 200~mK. This could be attributed primarily to an error in temperature probe measurements, which is beyond the scope of this work.

We thus conclude that calibration with an optimised set can achieve better results in terms of noise and absolute calibration using fewer calibrators and reduced calibration time, potentially minimizing possible drift errors in the system \citep{Bayesian_time_varying_Kirkham}.
Equivalently, maintaining a similar quality of calibration, this configuration also reduces the total noise in calibrated antenna temperature, as inferred analytically from Equation~\eqref{eq:final_Tant_noise}. 
Compared to the full-set case, the optimised calibrator set yields a similar value for the first term in the equation. Consequently, longer antenna observation becomes possible, reducing the second term which scales proportional to~$1/\tau_\mathrm{ant}$.



\section{Conclusions} \label{sec:conclusion}
REACH uses an extensive set of sources with widely varying temperatures and reflection coefficients for its receiver calibration. While the list of sources that can be used is not exhaustive, we have shown that an optimised set can improve calibration with fewer calibration sources. This also aids in reducing the period of a measurement spent in calibration, providing more time for the sky observation.
In this paper, we have introduced a method for such an optimisation of calibration sources based on condition numbers. The value of this function strongly depends on the matrix $\textbf{X}$, a crucial factor derived from the information of the calibrator set, which is used in the calibration equation (Equations~\ref{Eq:nwp_detailed_eq}--\ref{eq:all_cal_calibration_eq}). 

We started by establishing a relationship favouring a lower mean condition number for lower noise $\sigma_\mathrm{T}$ in the calibrated temperature solution. 
This was done by performing calibration of various validation sources in the REACH receiver using all possible combinations of calibrator sets. 
This knowledge was used for optimal choice of calibrators from the previously used pool of 12 sources in the REACH receiver. The optimised set selected was \texttt{[hot, r100, c2r36, c2r69, c10open, c10short]}, as obtained in Section~\ref{sec:best_set}. Through this, as previously expected, a diverse set of complex impedances (through resistance and cable length) and a heated thermal source were observed to be important to estimate the noise-waves accurately. 
This method was extended to a piecewise method where the `best' calibrator set is chosen for each frequency channel. This helped remove ripples in the condition number spectra, and therefore was expected to improve the calibration.

Finally, in Section~\ref{sec:calibration_results}, we presented the calibration results for all three methods of calibrator selection discussed in the paper, i.e. the full available set, the optimised set and piecewise calibrator optimisation. 
For a given dataset measured on the REACH radiometer, the source \texttt{c2r91} was calibrated, and the goodness of calibration was evaluated using noise or standard deviation $\sigma_\mathrm{T}$ and absolute mean deviation $\absval{\overline{\Delta T}}$ of the temperature solution.

To better compare the solutions with variable calibration time, we derived the quantity, total calibration time $\tau_\mathrm{cal}^\mathrm{total}$, that was used for normalising the noise in the solution. 
The solutions for the optimised calibrator set and the piecewise optimised set were shown to perform better than the full set for a similar total calibration time, with the full band optimised set showing an approximately 15~\% decrease in noise and an order of magnitude improvement in absolute calibration w.r.t. the full set.

With a reduced number of calibrators, we also reduced the calibration period, making the system less prone to drift errors and providing more time for sky observation, while maintaining the quality of the calibration. This optimisation becomes more important for space-based experiments like the CosmoCube, where drift error, volume and weight are key design constraints. 

Condition numbers serve as probes of numerical instabilities and systematics that may arise during the calibration process. This holds true for the method of conjugate priors calibration as well, which shall be investigated further \citep{CP_numerical_instability}.
While this tool was useful for the choice of calibration sources, it could also be used to develop better calibration strategies and design more efficient receiver systems for REACH and similar radiometer experiments.

\section*{Acknowledgements}
AKD was supported by the Cambridge Trust.
EdLA acknowledges the support of UKRI STFC via grant ST/V004425/1 (Ernest Rutherford Fellowship).
CJK was supported by the Science and Technology Facilities Council grant number ST/V506606/1.
The REACH collaboration acknowledges the support of its funders: UKRI (EP/Y02916X/1 - ERC COG Horizon Europe Guarantee), The Kavli Institute for Cosmology in Cambridge, the South African National Research Foundation, Stellenbosch University, and the \mbox{ALBORADA} Trust Fund.

\section*{Data Availability}

The data underlying this article will be shared on reasonable request to the corresponding author.



\bibliographystyle{mnras}
\bibliography{bibliography_akd} 




\appendix
\section{Uncertainty calculation for noise-dominated noise-wave parameters} 
\label{apdx_noise_dominated_nwp}

A measurement dataset in REACH consists of two parts, namely the calibration of the receiver and the sky measurement, where the receiver observes the antenna. Each calibrator observation is followed by observing a noise source and a reference load, which is used for Dicke switching. We denote the time observing each source during calibration as $\tau_\mathrm{cal}$. We need to calculate the uncertainty in the calibrated antenna temperature and its dependence on calibrator integration time. As the least-squares method is a frequency-by-frequency approach, so is the final relation of this derivation.

For the simplicity of this derivation, we have assumed that all calibrators have similar noise in their PSDs as they vary within an order, with minimal effects of impedance mismatches. We also assume no cross-correlations between the source powers.
The temperature probe measurements and the reflection coefficients are assumed noiseless, as their uncertainties are insignificant compared to those in PSDs. 
Another assumption is that all matrices in the further formulation that need to be inverted are perfectly invertible, with minimal numerical uncertainty, thus providing a condition number $\kappa~\approx~1$.

Then, the noise in power at the output of the receiver for each source can be shown as thermal noise. Using Equation~\eqref{eq:simple_power_gain}, we arrive at 
\begin{equation}
\begin{split}
    \sigma_\mathrm{cal}^2 &= \frac{b}{\tau_\mathrm{cal}} (T_\mathrm{cal}^2 + T_\mathrm{rec}^2) 
    \approx \sigma_\mathrm{L}^2 ,
    \\
    \sigma_\mathrm{NS}^2 &= \frac{b}{\tau_\mathrm{cal}} (T_\mathrm{NS}^2 + T_\mathrm{rec}^2) ,
\end{split}
\end{equation}
where $b$ is a proportionality constant which includes the gain of the receiver, and we take $T_\mathrm{cal} \approx T_\mathrm{L}$ 

As only the $X_\mathrm{NS}$ term will carry noise as it has the power terms, given 
\begin{equation*}
    X_\mathrm{NS} = \left( \frac{P_{\mathrm{cal}} - P_\mathrm{L}}{P_\mathrm{NS} - P_\mathrm{L}} \right)  X_\mathrm{L}  ,
\end{equation*}
the noise on this term is derived as 
\begin{equation} \label{app_eq:sigma^2_X_NS}
\begin{split}
    \sigma_{X_\mathrm{NS}}^2 &= X_\mathrm{NS}^2 
    \left[ 
    \frac{\sigma_\mathrm{cal}^2 + \sigma_\mathrm{L}^2}{(P_{\mathrm{cal}} - P_\mathrm{L})^2} 
    +
    \frac{\sigma_\mathrm{NS}^2 + \sigma_\mathrm{L}^2}{(P_{\mathrm{NS}} - P_\mathrm{L})^2}
    \right]
    \\
    &\approx  X_\mathrm{NS}^2 \frac{b}{\tau_\mathrm{cal}} d ,
\end{split}
\end{equation}
where 
\begin{equation*}
    d = \left[ 
    \frac{2 T_\mathrm{cal}^2 + 2 T_\mathrm{rec}^2 }{(P_{\mathrm{cal}} - P_\mathrm{L})^2} 
    +
    \frac{T_\mathrm{NS}^2 + T_\mathrm{L}^2 + 2T_\mathrm{rec}^2 }{(P_{\mathrm{NS}} - P_\mathrm{L})^2}
    \right]     .
\end{equation*}
The factor $d$ is assumed to be invariant over the calibrators as the terms are dominated by $P_\mathrm{NS}$, $T_\mathrm{NS}$, and $T_\mathrm{rec}$.
This noise $\sigma_{X_\mathrm{NS}}^2$ is useful to estimate the uncertainty in the noise-wave parameters estimation.

We first modify the calibration equation (Equation~\eqref{Eq:nwp_summing_equation}) as 
\begin{equation} \label{app_eq:Tcal_noise-noiseless_expansion}
\begin{split}
    T_\mathrm{cal} &= 
    \sum_{k \neq \mathrm{NS}} X_k \overline{\theta_k} 
    + X_\mathrm{NS} \overline{\theta_\mathrm{NS}}
    \\
    &= 
    \sum_{k \neq \mathrm{NS}} \overline{X_k} \overline{\theta_k}  
    + \overline{X_\mathrm{NS}} \overline{\theta_\mathrm{NS}} 
    + \delta X_\mathrm{NS} \overline{\theta_\mathrm{NS}}
          ,
\end{split}
\end{equation}
where $k \in \{ \mathrm{unc, cos, sin, NS, L} \}$ and shall denote an iteration variable for noise-wave parameters henceforth. 
An overbar denotes the noiseless component of the quantity.
We start with the central (or true) values of noise-wave parameters, which are equivalent to $\theta^\mathrm{fit}_\mathrm{NS}$ used in Section~\ref{Sec:Minimum_noise_estimation} for $\theta_\mathrm{NS}$ term. 
Similarly, $X_\mathrm{k} = \overline{X_\mathrm{k}} , \forall k \neq \mathrm{NS} $,
and the uncertainty in $X_\mathrm{NS}$ is denoted using the true value $\overline{X_\mathrm{NS}}$ and an error term $\delta X_\mathrm{NS}$, where
\begin{equation*}
\begin{split}
    X_\mathrm{NS} &= \overline{X_\mathrm{NS}} + \delta X_\mathrm{NS} ,
    \\
    \mathrm{Var}( \delta X_\mathrm{NS}) &= \sigma_{X_\mathrm{NS}}^2 ,
    \\ \text{and}
    \left \langle \delta X_\mathrm{NS} \right \rangle &= 0 .
\end{split}
\end{equation*}
We can separate the noiseless terms and the noise terms in Equation~\eqref{app_eq:Tcal_noise-noiseless_expansion} as
\begin{equation} \label{app_eq:Tcal+eta}
    T_\mathrm{cal} = \overline{T_\mathrm{cal}} + \eta_\mathrm{cal}  ,
\end{equation}
where 
\begin{equation}\label{app_eq:var_eta}
\begin{split}
    \eta_\mathrm{cal} &= \delta X_\mathrm{NS} \theta_\mathrm{NS} ,
    \\
    \mathrm{Var} (\eta_\mathrm{cal}) &= \theta_\mathrm{NS}^2  \sigma_{X_\mathrm{NS}}^2   ,
\end{split}
\end{equation}
and the covariance terms across calibrators $\mathrm{Cov}(\eta_i, \eta_j) = 0, (i \neq j)$ where $i$ and $j$ are calibrator indices. 

To derive the noise in the $\mathbf{\Theta}$ matrix, we first set up the least-squares formalism, where all X-terms are noiseless. For $N_\mathrm{cal}$ calibrators, the least-squares method for estimating the noise-wave parameters goes as follows.
\begin{equation} \label{app_eq:noiseless_lsq_minimization}
\begin{split}
    \mathrm{minimize} & \sum_{i=1}^{N_\mathrm{cal}} (T_i - \sum_k X_{ik} \theta_k)^2
    \\
    \implies &  
    \sum_{i=1}^{N_\mathrm{cal}} \mathbf{X}_i^T 
    (T_i - \mathbf{X}_i \mathbf{\Theta}) = 0
    \\
    \implies & 
    \hat{\mathbf{\Theta}} =
    \left(
    \sum_{i=1}^{N_\mathrm{cal}} \mathbf{X}_i^T \mathbf{X}_i 
    \right)^{-1}
    \left(
    \sum_{i=1}^{N_\mathrm{cal}} \mathbf{X}_i^T T_i 
    \right),
\end{split}
\end{equation}
where $X_{ik}$ is the $k^\mathrm{th}$ coefficient for noise-wave parameter for the $i^\mathrm{th}$ calibrator. Similarly, $\mathbf{X}_i$ is the row vector with coefficients corresponding to the $i^\mathrm{th}$ calibrator.

For the noisy $X_\mathrm{NS}$, following Equation~\eqref{app_eq:Tcal+eta}, the problem gets modified to
\begin{equation}
\begin{split}
    \mathrm{minimize} & \sum_{i=1}^{N_\mathrm{cal}} (T_i - \sum_k \overline{X_{ik}} \theta_k - \eta_i)^2
    \\
    \implies &  
    \sum_{i=1}^{N_\mathrm{cal}} \overline{\mathbf{X}_i}^T 
    (T_i - \overline{\mathbf{X}_i} \mathbf{\Theta} - \eta_i) = 0
    \\
    \implies & 
    \hat{\mathbf{\Theta}}_\mathrm{noisy} = 
    \left(
    \sum_{i=1}^{N_\mathrm{cal}} \overline{\mathbf{X}_i}^T \overline{\mathbf{X}_i}
    \right)^{-1}
    \left(
    \sum_{i=1}^{N_\mathrm{cal}} \overline{\mathbf{X}_i}^T (T_i-\eta_i) 
    \right)
    \\
    \implies &
    \hat{\mathbf{\Theta}}_\mathrm{noisy} =
    \hat{\mathbf{\Theta}} - 
    \left(
    \sum_{i=1}^{N_\mathrm{cal}} \overline{\mathbf{X}_i}^T \overline{\mathbf{X}_i}
    \right)^{-1}
    \left(
    \sum_{i=1}^{N_\mathrm{cal}} \overline{\mathbf{X}_i}^T \eta_i
    \right)   ,
\end{split}
\end{equation}
where $\hat{\mathbf{\Theta}}_\mathrm{noisy}$ is the noisy noise-wave parameter column matrix.
Then, we can calculate the covariance matrix $\mathbf{\Sigma_\Theta}$ for the noise-waves using Equations~\ref{app_eq:var_eta} and~\ref{app_eq:sigma^2_X_NS} as
\begin{equation}\label{app_eq:Sigma_Theta}
\begin{split}
    \mathbf{\Sigma_\Theta} &= 
    \left(
    \sum_{i=1}^{N_\mathrm{cal}} \overline{\mathbf{X}_i}^{T}\overline{\mathbf{X}_i}
    \right)^{-2}
    \left(\sum_{i=1}^{N_\mathrm{cal}} \overline{\mathbf{X}_i}^{T} \mathrm{Var}(\eta_i) \overline{\mathbf{X}_i}
    \right)
    \\
    &= 
    \theta_\mathrm{NS}^2  \sigma_{X_\mathrm{NS}}^2
    \left(
    \sum_{i=1}^{N_\mathrm{cal}} \overline{\mathbf{X}_i}^{T}\overline{\mathbf{X}_i}
    \right)^{-2}
    \left(
    \sum_{i=1}^{N_\mathrm{cal}} \overline{\mathbf{X}_i}^{T}\overline{\mathbf{X}_i}
    \right)
    \\
    & \approx
    \theta_\mathrm{NS}^2
    \frac{\overline{X_\mathrm{NS}}^2 b d}{\tau_\mathrm{cal}}
    \left(
    \sum_{i=1}^{N_\mathrm{cal}} \overline{\mathbf{X}_i}^{T}\overline{\mathbf{X}_i}
    \right)^{-1} .
\end{split}
\end{equation}

We note that the variance of the estimated noise-wave parameters varies inversely with the per-calibrator integration time. 
Moreover, as the calibrator number $N_\mathrm{cal}$ increases, the summation with quadratic terms increases monotonically, further reducing the variance in the noise-waves.
This, however, comes with the initial assumption that the matrix $\left(\sum_{i=1}^{N_\mathrm{cal}} \overline{\mathbf{X}_i}^{T}\overline{\mathbf{X}_i}\right)$ is perfectly invertible with a condition number close to one. In computation, this is not usually the case, and thus we require the calibrator set to be diverse with near-linearly independent columns and rows.

Finally, using the calibration equation in Equation~\eqref{Eq:nwp_summing_equation}, we can arrive at the noise in the final antenna (or validator) temperature with integration time $\tau_\mathrm{ant}$.
Similar to Equation~\eqref{app_eq:sigma^2_X_NS}, the noise in $X_\mathrm{ant, NS}$ for this case would be $\sigma_{X_\mathrm{ant, NS}}^2 = D / \tau_\mathrm{ant}$ for some constant $D$. 
Thus, following a similar calculation as for the calibrators and including the noise from the estimated noise-wave parameters, we arrive at
\begin{equation}
    T_\mathrm{ant} = \sum_k \overline{X_{\mathrm{ant}, k}} \theta_k + \delta X_{\mathrm{ant, NS}} \overline{\theta_\mathrm{NS}}    ,
\end{equation}
where we use noisy $\mathbf{\Theta}$, decompose $\theta_\mathrm{NS} = \overline{\theta_\mathrm{NS}} + \delta \theta_\mathrm{NS} $ and ignore the higher order noise terms. 
Then the variance in antenna temperature would be given by 
\begin{equation} \label{app_eq:final_Tant_noise}
    \sigma_\mathrm{ant}^2 = 
    \mathbf{X}_\mathrm{ant} \mathbf{\Sigma_\Theta} \mathbf{X}_\mathrm{ant}^T 
    + \overline{\theta_\mathrm{NS}}^2 \frac{D}{\tau_\mathrm{ant}}   .
\end{equation}

We observe that for noisy noise-wave parameters, within the least-squares formalism, the antenna temperature uncertainty does not follow the radiometric equation, but rather is a weighted sum of contributions from both calibrator noise carried into noise-waves (first term) and the noise in antenna observation (second term).
In a case where the noise-waves are fitted with functions without noise, e.g. using conjugate priors \citep{Bayesian_nwp_conjugate_priors}, it is observed that $\sigma^2_\mathrm{ant}$ scaled inversely with antenna integration time $\tau_\mathrm{ant}$, following the radiometric equation.
Similarly, in a case where antenna observation is noiseless, an increase in calibrator number $N_\mathrm{cal}$ or per-calibrator integration time $\tau_\mathrm{cal}$ would significantly reduce $\sigma_\mathrm{ant}$.


\bsp	
\label{lastpage}
\end{document}